\documentclass[aps,floats]{revtex4}
\usepackage{amsmath,amssymb}
\usepackage{graphicx,epsfig}
\usepackage[greek,english]{babel}
\usepackage{bbold}

\begin{document}
\bibliographystyle {plain}
\pdfoutput=1
\def\oppropto{\mathop{\propto}} 
\def\opsimeq{\mathop{\simeq}}
\def\opoverderline{\mathop{\overline}}
\def\operarrow{\mathop{\longrightarrow}}
\def\opsim{\mathop{\sim}}

\def\opmin{\mathop{\min}} 
\def\opmax{\mathop{\max}} 
\def\oplim{\mathop{\lim}}

\def\fig#1#2{\includegraphics[height=#1]{#2}}
\def\figx#1#2{\includegraphics[width=#1]{#2}}

\title{  
Large deviations for the Skew-Detailed-Balance Lifted-Markov processes \\
to sample the equilibrium distribution of the Curie-Weiss model
  } 


\author{ C\'ecile Monthus }
 \affiliation{Institut de Physique Th\'{e}orique, 
Universit\'e Paris Saclay, CNRS, CEA,
91191 Gif-sur-Yvette, France}

\begin{abstract}

Among the Markov chains breaking detailed-balance that have been proposed in the field of Monte-Carlo sampling in order to accelerate the convergence towards the steady state with respect to the detailed-balance dynamics,  the idea of 'Lifting' consists in duplicating the configuration space into two copies $\sigma=\pm$ and in imposing directed flows in each copy in order to explore the configuration space more efficiently. The skew-detailed-balance Lifted-Markov-chain introduced by K. S. Turitsyn, M. Chertkov and M. Vucelja [Physica D Nonlinear Phenomena 240 , 410 (2011)] is revisited for the Curie-Weiss mean-field ferromagnetic model, where the dynamics for the magnetization is closed. The large deviations at various levels for empirical time-averaged observables are analyzed and compared with their detailed-balance counterparts, both for the discrete extensive magnetization $M$ and for the continuous intensive magnetization $m=\frac{M}{N}$ for large system-size $N$.

\end{abstract}

\maketitle


\section{ Introduction }

In the field of Monte-Carlo algorithms,
the goal is to sample the equilibrium distribution $ P_{eq}(C)$ over configurations $C$ 
via the dynamical trajectory $C(0 \leq t \leq T) $ of some Markov chain
converging to $ P_{eq}(C)$  (see the book \cite{werner_book} and references therein),
so that the empirical time-average of any observable $O(C)$  
will converge for large $T$ towards
the average of $O(C)$ over the equilibrium distribution $P_{eq}(C)$
\begin{eqnarray}
\frac{1}{T} \int_0^{T} dt O (  C(t) )  
\opsimeq_{T \to \infty}
 \sum_{ C} O (C) P_{eq}(C)
\label{ergodic}
\end{eqnarray} 
The first natural possibility is to choose a Markov chain satisfying Detailed-Balance, with no current in the steady state, so that the Markov chain is a faithful 
equilibrium dynamics, with the drawback of possible slow convergence properties, in particular at criticality
with the famous phenomenon of "critical slowing down", or whenever dynamical barriers are difficult to overcome. As a consequence, it is interesting to consider 
instead Markov chains breaking Detailed-Balance, with currents in the steady state,
so that the Markov chain is not a faithful equilibrium dynamics, and can thus converge faster
towards the equilibrium distribution $P_{eq}(C) $. 
Among the various proposals of Markov chains breaking Detailed-Balance
in the field of Monte-Carlo sampling, the idea of 'Lifting'
\cite{proc,diaconis,TCV,weigel,SDBIsing1d,beyondDB,vuceljaa,SDBeigen,SDBcurierun,SDBpotts,Werner_review} consists in duplicating the configuration space into two copies $\sigma=\pm$,
in order to introduce steady currents in each copy to move more efficiently, while the switching events between the two copies will ensure the convergence towards $P_{eq}(C)$
after summing over $\sigma=\pm 1$.
In particular, Konstantin Turitsyn, Michael Chertkov and Marija Vucelja
 \cite{TCV}
have introduced the notion of Skew-Detailed-Balance to choose the rates within each copy,
and the choice of fully irreversible switching flows between the two copies $\sigma=\pm 1$,
i.e. for each configuration $C$, one of the two switching rates vanishes.
This algorithm has been first applied to the Curie-Weiss mean-field ferromagnetic model
\cite{TCV,weigel,vuceljaa,SDBcurierun}
in order to analyze the acceleration of convergence with respect to the detailed-balance algorithm :
the main conclusion is that the convergence is faster both in the high-temperature phase $\beta<\beta_c$ and at criticality
$\beta=\beta_c$, with the following scaling with respect to the system-size $N$
of the ratio between the characteristic equilibrium times $\tau^{SDB}_{eq} $ and $\tau^{DB}_{eq} $ with Skew-Detailed-Balance and with Detailed-Balance
respectively
\cite{TCV,weigel,vuceljaa,SDBcurierun}
 \begin{eqnarray}
\frac{ \tau^{SDB}_{eq} }{\tau^{DB}_{eq} } \oppropto_{N \to +\infty}  \frac{1}{ N^{\frac{1}{2}}  } \ \ \ \ \ {\rm for } \ \ \beta < \beta_c 
\nonumber \\
\frac{ \tau^{SDB}_{eq} }{\tau^{DB}_{eq} } \oppropto_{N \to +\infty}  \frac{1}{ N^{\frac{3}{4}}  } \ \ \ \ \ {\rm at } \ \ \beta= \beta_c 
\label{ratiotaueqSDB}
\end{eqnarray}
Applications of this algorithm to other spin models can be found in \cite{weigel,SDBIsing1d,beyondDB,SDBpotts} 
and are reviewed in \cite{vuceljaa}.

 As proposed in the context of various non-reversible algorithms
 \cite{rey_langevin,rey_graph,rey_general,jack_accel,chetrite_sampling},
 it is interesting to characterize their acceleration of convergence 
 via  the recent progress concerning the large deviation properties
   of empirical time-averaged observables for non-equilibrium steady-states.
  Within the theory of large deviations  
(see the reviews \cite{oono,ellis,review_touchette} and references therein),
the traditional classification for dynamical trajectories 
over a large time-window $T$ \cite{oono,review_touchette}
involves three nested levels : 
 the Level 1 concerns the empirical time-averaged value of 
 observables $O$ (i.e. the left hand side of Eq. \ref{ergodic});  
the Level 2 concerns the empirical time-averaged density $\rho_T(C)$,
i.e. the histogram of the configurations visited during $T$
that will converge towards the steady-state $P_{eq}(C)$ for large $T$,
while the Level 3 characterizes the whole empirical process.
For non-equilibrium steady states with non-vanishing steady currents,
this classification has turned out to be inadequate,
because the Level 2 is not closed, while the Level 3 is too general.
The introduction of the Level 2.5 
concerning the joint probability distribution of the empirical time-averaged density 
and of the empirical time-averaged flows has thus been a major achievement.
 Indeed, the rate functions at Level 2.5 are explicit for various types of Markov processes,
including discrete-time Markov Chains
 \cite{fortelle_thesis,fortelle_chain,review_touchette,c_largedevdisorder,c_reset,c_inference},
continuous-time Markov Jump processes
\cite{fortelle_thesis,fortelle_jump,maes_canonical,maes_onandbeyond,wynants_thesis,chetrite_formal,BFG1,BFG2,chetrite_HDR,c_ring,c_interactions,c_open,c_detailed,barato_periodic,chetrite_periodic,c_reset,c_inference,c_runandtumble,c_jumpdiff}
and Diffusion processes described by Fokker-Planck equations
\cite{wynants_thesis,maes_diffusion,chetrite_formal,engel,chetrite_HDR,c_reset,c_lyapunov,c_inference}.
As a consequence, the explicit Level 2.5 can be considered as a starting point
 for any Markov process converging towards some steady state, 
 from which many other large deviations properties can be derived
via contraction. In particular, the Level 2 for the empirical density alone 
can be obtained via the optimization of the Level 2.5 over the empirical flows,
while the Level 1 can be obtained from the Level 2 by the further appropriate contraction.
More generally, the Level 2.5 can be contracted
to obtain the large deviations properties of any time-additive observable
of the dynamical trajectory involving both the configuration and the flows.
The link with the studies of general time-additive observables 
via deformed Markov operators  \cite{derrida-lecture,sollich_review,lazarescu_companion,lazarescu_generic,jack_review,vivien_thesis,lecomte_chaotic,lecomte_thermo,lecomte_formalism,lecomte_glass,kristina1,kristina2,jack_ensemble,simon1,simon2,simon3,Gunter1,Gunter2,Gunter3,Gunter4,chetrite_canonical,chetrite_conditioned,chetrite_optimal,chetrite_HDR,touchette_circle,touchette_langevin,touchette_occ,touchette_occupation,derrida-conditioned,derrida-ring,bertin-conditioned,touchette-reflected,touchette-reflectedbis,c_lyapunov,previousquantum2.5doob,quantum2.5doob,quantum2.5dooblong,c_ruelle}
 can be then understood via the corresponding 'conditioned' process obtained from the generalization of Doob's h-transform.
 
 
 In this paper, the Skew-Detailed-Balance Lifted-Markov processes for the Curie-Weiss model
  \cite{TCV,weigel,SDBcurierun,vuceljaa} are revisited
  to analyze their large deviations at various levels
  and to compare them with their counterparts
for the Detailed-Balance dynamics.  


The paper is organized as follows.
The links between the various levels of large deviations
 for Markov processes are summarized in section \ref{sec_largedev}.
 This formalism is then applied to compare various Markov dynamics for 
 the Curie-Weiss model, whose equilibrium properties are briefly recalled in section \ref{sec_eq}.
The single-spin-flip dynamics is analyzed for the discrete extensive magnetization $M$
both for the standard Detailed-Balance Markov chain
in section \ref{sec_DBjump}, and for the Skew-Detailed-Balance Lifted-Markov-Chain
in section \ref{sec_SkewDBjump}.
Then we focus on the continuous  intensive magnetization $m$
in order to analyze the corresponding effective dynamics, 
namely the Detailed-Balance diffusion process in section \ref{sec_DBdiff},
and the Skew-Detailed-Balance run-and-tumble process
in section \ref{sec_SkewDBrun}.
Our conclusion are summarized in section \ref{sec_conclusion}.
Appendix A describes the properties of the Green function $G$ governing the typical convergence properties.
The next four appendices contain the explicit computations
 of the Green functions for the four dynamics considered in the main text.


\section{ Reminder on large deviations at various levels for Markov processes   }

\label{sec_largedev}

In this section, we recall the links between the various levels of large deviations
for the case of continuous-time Markov chain in discrete configuration space.

\subsection{ Continuous-time Markov Chain of generator $w$ with its zero-eigenvalue subspace}

The Markov jump process of generator $w$
\begin{eqnarray}
 \partial_t P_{t}( x )    =   \sum_{y }  w(x,y) P_t(y)
\label{master}
\end{eqnarray}
is assumed to converge towards the steady state $P_*(x) $ satisfying
\begin{eqnarray}
 0   =   \sum_{y }  w(x,y) P_*(y)
\label{mastersteady}
\end{eqnarray}
The off-diagonal $x \ne y$ matrix element $w(x,y) \geq 0$   
represents the transition rate from $y$ to $x $,
while the diagonal elements are negative $w(x,x) \leq 0 $ and are fixed by the conservation of probability to be
\begin{eqnarray}
w(x,x) \equiv  - \sum_{y \ne x} w(y,x) 
\label{wdiag}
\end{eqnarray}
Eqs  \ref{mastersteady} and \ref{wdiag}  mean that 
 the highest eigenvalue of the Markov matrix $w(.,.)$ is zero 
 \begin{eqnarray}
 0 && = \langle l_0 \vert w = \sum_x l_0(x) w(x,y)
 \nonumber \\
 0  && =  w \vert r_0 \rangle =  \sum_{y }  w(x,y) r_0(y)
\label{mastereigen0}
\end{eqnarray}
where the positive left eigenvector is trivial 
\begin{eqnarray}
 l_0(x)=1
\label{markovleft}
\end{eqnarray}
while the positive right eigenvector is given by the steady state
\begin{eqnarray}
 r_0(y)=P_*(y) 
\label{markovright}
\end{eqnarray}
with the normalization
\begin{eqnarray}
1 = \langle l_0 \vert r_0 \rangle = \sum_x l_0(x) r_0(x)
\label{markovrightleftnorma}
\end{eqnarray}

\subsection{ Large deviations at Level 1 for the empirical time-averaged observable ${\cal O}_T $  }

For a long trajectory $x(0 \leq t \leq T) $ of the Markov jump process,
 we are interested into the empirical time-averaged observable of the position $x(t)$ involving the observable $O(x) $
\begin{eqnarray}
{\cal O}_T \equiv \frac{1}{T} \int_0^{T} dt  O \left(  x(t)  \right) 
\label{defadditive}
\end{eqnarray} 
The probability distribution
to see the value ${\cal O}_T=O$
follows the large deviation form with respect to $T$
  \begin{eqnarray}
 P^{[Level1]}_T[ O ] \opsimeq_{T \to +\infty} e^{- T I_1 [ O ]}
\label{level1def}
\end{eqnarray} 
where the rate function $I_1 [ O ] \geq 0 $ at Level 1 vanishes only for the steady value 
\begin{eqnarray}
O_* \equiv \sum_x O(x) P_*(x) 
\label{aeq}
\end{eqnarray} 
Equivalently, the generating function
\begin{eqnarray}
\langle e^{k T {\cal O}_T } \rangle \equiv \int dO  e^{k T O} P^{[Level1]}_T[ O ] \opsimeq_{T \to +\infty} 
\int dO e^{ T \left[ k O- I_1 [ O ] \right] }\opsimeq_{T \to +\infty} e^{ T \mu(k) }
\label{level1gen}
\end{eqnarray} 
can be evaluated via the saddle-point method in $T$ to obtain that 
the scaled cumulant generating function $\mu(k)$ is the Legendre transform of the rate function $ I_1 [ O ] $.


\subsubsection{ Scaled cumulant generating function $\mu(k)$ as the highest eigenvalue of some deformed matrix}

The standard method to analyze time-additive observables of stochastic processes
goes back to the famous Feynman-Kac formula for diffusion processes
and consists in studying the appropriate 'tilted' dynamical process.
This approach based on deformed Markov generators has been used extensively in the field of non-equilibrium processes recently
 \cite{derrida-lecture,sollich_review,lazarescu_companion,lazarescu_generic,jack_review,vivien_thesis,lecomte_chaotic,lecomte_thermo,lecomte_formalism,lecomte_glass,kristina1,kristina2,jack_ensemble,simon1,simon2,simon3,Gunter1,Gunter2,Gunter3,Gunter4,chetrite_canonical,chetrite_conditioned,chetrite_optimal,chetrite_HDR,touchette_circle,touchette_langevin,touchette_occ,touchette_occupation,derrida-conditioned,derrida-ring,bertin-conditioned,touchette-reflected,touchette-reflectedbis,c_lyapunov,previousquantum2.5doob,quantum2.5doob,quantum2.5dooblong,c_ruelle}.
For the Markov jump dynamics of Eq. \ref{master}, one can use
the probabilities of dynamical trajectories
\begin{eqnarray}
{\cal P}[x(0 \leq t \leq T)]   
=  e^{ \displaystyle  \left[  \sum_{t: x(t^+) \ne x(t^-) } \ln ( w(x(t^+) , x(t^-)  ) +  \int_0^T dt  w(x(t) , x(t) )  \right]  }
\label{pwtrajjump}
\end{eqnarray}
to compute the generating function of Eq. \ref{level1gen} 
\begin{eqnarray}
\langle e^{k T {\cal O}_T } \rangle \equiv \langle e^{k \int_0^T dt O(x(t)) } \rangle
\label{level1gend}
\end{eqnarray} 
One obtains that $\mu(k)$ is the highest eigenvalue of the deformed matrix
\begin{eqnarray}
w^{[k]}(x,y) = w(x,y) +\delta_{x,y} k O(x)
\label{wdeformed}
\end{eqnarray} 
where only the diagonal elements are changed into
\begin{eqnarray}
w^{[k]}(x,x) = w(x,x) + k O(x) = - \sum_{x \ne y} w(y,x) + k O(x)
\label{wdeformeddiag}
\end{eqnarray}


\subsubsection{ First cumulants via the perturbation theory in $k$}

\label{subsec_per}

The perturbation theory in $k$ for the highest eigenvalue
\begin{eqnarray}
\mu(k) = 0 +  \mu_1 k +k^2 \mu_2 + \mu_3 k^3 +\mu_4 k^4 +O(k^5)
\label{pereigenjump}
\end{eqnarray}
of the deformed operator of Eq. \ref{wdeformed} involving the perturbation
 \begin{eqnarray}
w_{per}(x,y) \equiv \delta_{x,y}  O(x)
\label{wper}
\end{eqnarray} 
 is 
very similar to the standard perturbation theory of quantum mechanics,
except for the existence of left and right eigenvectors
(see for instance the reminder in Appendix A of \cite{c_ruelle}):

(i) The first-order correction to the eigenvalue of Eq. \ref{pereigenjump}
corresponds to the averaged value of the perturbation $w_{per} $
 computed in the unperturbed zero-eigenvalue subspace
\begin{eqnarray}
\mu_1 = \langle  l_0 \vert  w_{per}  \vert r_0 \rangle =  \sum_x O(x) P_*(x) = O_*
  \label{energy1}
\end{eqnarray}
and corresponds to Eq. \ref{ergodic} as it should.

(ii) The second-order correction to the eigenvalue
involves the Green function $G$ of Eq. \ref{Gasinverse}, whose properties are recalled in detail in Appendix \ref{sec_green}
\begin{eqnarray}
\mu_2  = 
 \langle  l_0 \vert  w_{per}  G w_{per}  \vert r_0 \rangle
  =  \sum_x O(x) \sum_y O(y) G(x,y) P_*(y)
  \label{energy2r}
\end{eqnarray}

(iii) The Green function $G$ also governed the higher cumulants,
for instance the third cumulant via the third-order correction
\begin{eqnarray}
\mu_3 = 
 \langle  l_0 \vert  w_{per}  G w_{per} G w_{per} \vert r_0 \rangle
  - \mu_1 \langle  l_0 \vert  w_{per}  G^2 w_{per}  \vert r_0 \rangle
  \label{energy3r}
\end{eqnarray}
and the fourth cumulant via 
the fourth-order correction
\begin{eqnarray}
\mu_4 && = 
 \langle  l_0 \vert  w_{per}  G w_{per} G w_{per} G w_{per}\vert r_0 \rangle
  - \mu_2 \langle  l_0 \vert  w_{per}  G^2 w_{per}  \vert r_0 \rangle
\nonumber \\
&&    -\mu_1 \langle  l_0 \vert  w_{per}  G^2 w_{per} G w_{per} \vert r_0 \rangle
     -\mu_1 \langle  l_0 \vert  w_{per}  G w_{per} G^2 w_{per} \vert r_0 \rangle
     +  \mu_1^2 \langle  l_0 \vert  w_{per}  G^3 w_{per}  \vert r_0 \rangle
  \label{energy4r}
\end{eqnarray}


\subsubsection{ Discussion}

In conclusion, if one wishes to obtain the full large deviations properties
of the empirical time-averaged observable ${\cal O}_T $, 
one needs to compute the highest eigenvalue $\mu(k)$ of the deformed operator of Eq. \ref{wdeformed}
for any $k$.
While this is not possible in many interesting models, 
one can always consider the perturbative expansion around $k=0$ mentioned above,
or one can instead start from the explicit large deviations at Level 2.5 as explained in the next
subsection.


\subsection{ Explicit large deviations at Level 2.5 for the time-averaged density and the time-averaged flows  }

\label{app_2.5}

As recalled in the Introduction, the large deviations at Level 2.5
are explicit for various types of Markov processes.
For a very long dynamical trajectory $x(0 \leq t \leq T)$
of the Markov jump process of Eq. \ref{master},
the empirical time-averaged density $\rho_T( x ) $
measures the histogram of the various configurations $x$
seen during the dynamical trajectory $x(0 \leq t \leq T) $ 
\begin{eqnarray}
 \rho_T( x ) \equiv \frac{1}{T} \int_0^T dt \ \delta_{x(t) , x }
\label{rhoc}
\end{eqnarray}
while the empirical time-averaged flows characterize the jumps 
seen during the dynamical trajectory $x(0 \leq t \leq T) $
\begin{eqnarray}
q_T(x,y) \equiv  \frac{1}{T} \sum_{t : x(t^+) \ne x(t^-)} \delta_{x(t^+),x} \delta_{x(t^-),y} 
\label{jumpempiricaldensity}
\end{eqnarray}
The joint probability to see the empirical density
 $\rho( x )$ 
and to see the empirical flows $q(x,y) $
follows the
large deviation form  
\cite{fortelle_thesis,fortelle_jump,maes_canonical,maes_onandbeyond,wynants_thesis,chetrite_formal,BFG1,BFG2,chetrite_HDR,c_ring,c_interactions,c_open,c_detailed,barato_periodic,chetrite_periodic,c_reset,c_inference,c_runandtumble,c_jumpdiff}
\begin{eqnarray}
P^{Level \ 2.5}_{T}[ \rho(.) ; q(.,.) ] \oppropto_{T \to +\infty} \delta \left( \sum_x \rho(x) - 1 \right) 
 \prod_x \left[ \sum_{y \ne x} \left( q(x,y) - q(y,x) \right) \right] e^{- T I_{2.5}[ \rho(.) ; q(.,.) ] }
\label{level2.5}
\end{eqnarray}
that involves the normalization constraint for $\rho(.)$
and the stationarity constraints for the flows $ q(.,.)$,
i.e. for any $x$, the total incoming flow into $x$ should be equal 
to the total outgoing flow out of $x$.
The rate function at level 2.5 is explicit
for any continuous-time Markov chain generator $w$
\begin{eqnarray}
I_{2.5}[ \rho(.) ; q(.) ]=  \sum_{y } \sum_{x \ne y} 
\left[ q(x,y)  \ln \left( \frac{ q(x,y)  }{  w(x,y)  \rho(y) }  \right) 
 - q(x,y)  + w(x,y)  \rho(y)  \right]
\label{rate2.5}
\end{eqnarray}


\subsection{ Contraction towards the Level 2 for the empirical density $\rho(.)$ alone and towards the Level 1 }

The large deviations at Level 2
for the probability of the empirical density $\rho(.)$ alone 
can be obtained via integration over the emprical flow $q(.,.)$
of the joint probability at Level 2.5 of Eq. \ref{level2.5}
\begin{eqnarray}
P^{Level \ 2}_{T}[ \rho(.)  ] && \oppropto_{T \to +\infty} \delta \left( \sum_x \rho(x) - 1 \right) 
\int {\cal D} q (.,.) 
 \prod_x \left[ \sum_{y \ne x} \left( q(x,y) - q(y,x) \right) \right] e^{- T I_{2.5}[ \rho(.) ; q(.,.) ] }
\nonumber \\
&&  \oppropto_{T \to +\infty} \delta \left( \sum_x \rho(x) - 1 \right) e^{- T I_{2}[ \rho(.)  ] }
\label{level2}
\end{eqnarray}
So the rate function $ I_{2}[ \rho(.)  ] $ at Level 2 will be explicit only if one is able to 
find the optimal flows $q^{opt}(.,.)$ that dominate the integral over $q(.,.)$ above,
as a function of the given empirical density $\rho(.)$. 

If the Level 2 is explicit, one can try to compute the rate function at Level 1 of Eq. \ref{level1def}
via the contraction of the Level 2 of Eq. \ref{level2}
as follows
 \begin{eqnarray}
 P^{[Level1]}_T[ O ]  \opsimeq_{T \to +\infty} \int {\cal D} \rho(.) 
 \delta \left( O -  \sum_x O(x)   \rho(x) \right)
 \delta \left( \sum_x \rho(x) - 1 \right) e^{- T I_{2}[ \rho(.)  ] }
\label{level12}
\end{eqnarray} 
Equivalently, the generating function of Eq. \ref{level1gen} corresponds to the contraction
\begin{eqnarray}
\langle e^{k T {\cal O}_T } \rangle \equiv 
 \opsimeq_{T \to +\infty} \int {\cal D} \rho(.) 
 \delta \left( \sum_x \rho(x) - 1 \right) e^{ T \left( k \sum_x O(x)   \rho(x)- I_{2}[ \rho(.)  ] \right) }
\label{level1genfrom2}
\end{eqnarray}


\subsection{ Diffusion processes in continuous space in dimension $d$ : explicit large deviations at level 2.5  }

\label{subsec_2.5diff}

For the Fokker-Planck dynamics 
in dimension $d$
involving the diffusion coefficient $D(\vec x)$ and the force field $\vec F(\vec x)$
\begin{eqnarray}
 \partial_t P_{t}( \vec x )  
   =   \partial_{\vec x} \left[ - \vec F(\vec x) P_{t}( \vec x ) + D(\vec x)  \partial_{\vec x} P_{t}( \vec x ) \right]
\label{fokkerplanck}
\end{eqnarray}
the joint distribution of the empirical density $\rho(.)$ and of the empirical current $\vec j(\vec x)$
satisfy the large deviation form \cite{wynants_thesis,maes_diffusion,chetrite_formal,engel,chetrite_HDR,c_reset,c_lyapunov,c_inference}
\begin{eqnarray}
 P_T[ \rho(.), \vec j(.)]   \opsimeq_{T \to +\infty}  \delta \left(\int d^d \vec x \rho(\vec x) -1  \right)
\left[ \prod_{\vec x }  \delta \left(  \vec \nabla . \vec j(\vec x) \right) \right] 
e^{- \displaystyle T I_{2.5} [ \rho(.), \vec j(.)]
 }
\label{ld2.5diff}
\end{eqnarray}
that involves the normalization constraint for the density$\rho(.)$
and the divergence-free stationarity constraint for the current $ \vec j(\vec x)$,
while the rate function at level 2.5 reads
\begin{eqnarray}
I_{2.5} [ \rho(.), \vec j(.)]  =
\int \frac{d^d \vec x}{ 4 D (\vec x) \rho(\vec x) } \left[ \vec j(\vec x) - \rho(\vec x) \vec F(\vec x)+D (\vec x) \partial_{\vec x} \rho(\vec x) \right]^2
\label{rate2.5diff}
\end{eqnarray}


\section{ Reminder on the thermal equilibrium of the Curie-Weiss model  }

\label{sec_eq}

In the following sections, we will analyzed in detail various Markov dynamics 
that converge towards the thermal equilibrium of the Curie-Weiss model,
whose properties are briefly recalled in the present section.

\subsection{ Curie-Weiss model : mean-field ferromagnetic model for $N$ spins with $2^N$ configurations }

The Curie-Weiss model is the mean-field ferromagnetic model of coupling $J>0$ for
$N$ spins $S_i=\pm$ with $i=1,2,..,N$, with $2^N$ configurations $C=\{S_1,...,S_N\}$ of energy
\begin{eqnarray}
{\cal E}(C=\{S_1,...,S_N\}) \equiv - \frac{J}{2 N }  \sum_{i=1}^N \sum_{j=1}^N  S_i S_j 
\label{ec}
\end{eqnarray}
At inverse temperature $\beta$, the probability to see the configuration $C$ follows the Boltzmann distribution
\begin{eqnarray}
 {\cal P}^{eq}_N(C) &&= \frac{ e^{-\beta {\cal E}(C)} }{Z_N}
 \nonumber \\
 Z_N && = \sum_{C} e^{-\beta {\cal E}(C)}
\label{Pconfig}
\end{eqnarray}
In the two following subsections, 
we recall the equilibrium distributions of the extensive magnetization $M$ for finite size $N$,
and of the intensive magnetization $m$ for large $N \to + \infty$.


\subsection{ Equilibrium distribution for the even extensive magnetization $M  \in \{-N,..,-2,0,2,...,N\}$  }

For definiteness, we will only consider the case of an even number $N$ of spins, 
where the extensive magnetization
\begin{eqnarray}
M \equiv  \sum_{i=1}^N  S_i 
\label{magn}
\end{eqnarray}
takes the $(N+1)$ even values
\begin{eqnarray}
M   \in \{-N,-N+2,..,-2,0,2,...,N-2,N\}
\label{even}
\end{eqnarray}
The energy of the configuration $C=\{S_1,...,S_N\}$ in Eq. \ref{ec}
only depends on its magnetization $M$ 
\begin{eqnarray}
{\cal E}(C=\{S_1,...,S_N\}) 
=  -  \frac{J }{2 N } \left( \sum_{i=1}^N   S_i \right)^2
= -  \frac{J }{2 N } M^2 
\label{HN}
\end{eqnarray}
 or equivalently of the numbers $N_{\pm}$ of positive and negative spins among the $N=N_+ + N_-$ spins 
\begin{eqnarray}
N_+ && =  \frac{N+M}{2}
\nonumber \\
N_- && =  \frac{N-M}{2}
\label{Npm}
\end{eqnarray}
The number $\Omega(M)$ of configurations with magnetization $M$ is given by the binomial coefficient
\begin{eqnarray}
\Omega_N(M) = \frac{ N!}{ N_+! N_-!}
=\frac{ N!}{ \left(  \frac{N+M}{2}\right)! \left( \frac{N-M}{2} \right)!} 
\label{omega}
\end{eqnarray}
As a consequence, the thermal equilibrium of Eq. \ref{Pconfig} for the $2^N$ configurations
can be projected onto the probability to see the even magnetization $M$ 
\begin{eqnarray}
P^{eq}_N(M) = \frac{\Omega_N(M) e^{  \frac{\beta J }{2 N } M^2} }{Z_N} 
\label{Pnpeq}
\end{eqnarray}
with the normalization
\begin{eqnarray}
Z_N = \sum_{M=-N}^N 
 \Omega_N(M) e^{  \frac{\beta J }{2 N } M^2}
\label{PMbig}
\end{eqnarray}
{\it Note that in the whole paper, in any sum concerning the extensive magnetization $M$, it will be implied that $M$ is even.}

The convergence towards the equilibrium distribution of Eq. \ref{Pconfig} and Eq. \ref{Pnpeq} 
will be analyzed for the single-spin-flip Markov dynamics
either with detailed-balance in section \ref{sec_DBjump}
or with the skew-detailed-balance in section \ref{sec_SkewDBjump}.


\subsection{ Large deviations for the intensive magnetization $m=\frac{M}{N} \in [-1,+1]$ for large size $N \to +\infty$  }

For large size $N\to +\infty $, one wishes to focus on the intensive magnetization 
\begin{eqnarray}
m \equiv \frac{M}{N} = \frac{1}{N}  \sum_{i=1}^N  S_i  \in [-1,+1]
\label{magni}
\end{eqnarray}
Using the Stirling approximation for the factorials in Eq. \ref{omega}
\begin{eqnarray}
\Omega_N(M=Nm) \opsimeq_{N \to +\infty} \frac{2^N}{\sqrt{ N \frac{1-m^2}{2} }}
e^{ - N \left[ \frac{(1+m) }{2} \ln (1+m) + \frac{(1-m)}{2}  \ln (1-m) \right] }
\label{omegastirling}
\end{eqnarray}
one obtains that the probability $p^{eq}_N(m) $ to see the intensive magnetization $m$
follows the large deviation form for large size $N$
\begin{eqnarray}
  p^{eq}_N(m) \opsimeq_{N \to +\infty} 
 e^{-  N  \left[ V(m) - V_{min} \right] } 
\label{Psmall}
\end{eqnarray}
with the effective potential 
\begin{eqnarray}
 V(m) = \frac{(1+m) }{2} \ln (1+m) + \frac{(1-m)}{2}  \ln (1-m)   - \frac{\beta J  }{2} m^2
\label{vm}
\end{eqnarray}
while $V_{min}$ is the minimum value of this potential.
The two first derivatives of the effective potential of Eq. \ref{vm}
\begin{eqnarray}
 V'(m) && = \frac{ \ln (1+m) - \ln (1-m) }{2}    - \beta J m
 \nonumber \\
  V''(m) && = \frac{ \frac{1}{1+m}  +  \frac{1}{1-m} } {2}- \beta J  =  \frac{1}{1-m^2} - \beta J 
\label{vmderi}
\end{eqnarray}
and the perturbative expansion around zero magnetization $m=0$
\begin{eqnarray}
 V(m) = (1  - \beta J ) \frac{m^2 }{2} + \frac{m^4}{12} +O(m^6)
\label{vmseries}
\end{eqnarray}
allows to identify the ferromagnetic transition at the critical point $ \beta_c J=1$ :

(i) in the high temperature phase $\beta J < \beta_c J=1$ : 
the curvature of the potential is always positive $V''(m)>0$,
the derivative $V'(m)$ is negative for $m<0$, vanishes at $m=0$ and is positive for $m>0$.
So the minimum $V_{min}$ of the potential is at zero magnetization $m=0$ 
\begin{eqnarray}
V_{min}= 0 = V(m=0)
\label{vminzero}
\end{eqnarray}
with the positive curvature $V''(m=0)=1-\beta J >0$.
For later purposes, it is useful to write explicitly the sign of the derivative of the potential in this high temperature phase $\beta J < \beta_c J=1$
\begin{eqnarray}
&& V'(m) <0 \ \ \ \ {\rm for } \ \ -1<m< 0
 \nonumber \\
&& V'(m) >0 \ \ \ \ {\rm for } \ \ 0<m< 1
\label{vprimehigh}
\end{eqnarray}

(ii) at the critical point $ \beta_c J=1$, the minimum of the potential is still
at $m=0$ (Eq. \ref{vminzero}) but the corresponding curvature vanishes $V''(0)=1-\beta_c J =0$,
so that the leading term in the series expansion of Eq. \ref{vmseries} is of order $m^4$.

(iii) in the low temperature phase $\beta J < \beta_cJ =1$, $m=0$ is a local maximum,
while the two symmetric minima at $(\pm m_{sp}(\beta)) \ne 0 $ satisfying $V'(\pm m_{sp}(\beta))=0 $
represent the spontaneous magnetization of ferromagnetic phase.
For later purposes, it is useful to write explicitly the sign of the derivative of the potential in this low temperature phase $\beta J > \beta_c J=1$
\begin{eqnarray}
&& V'(m) <0 \ \ \ \ {\rm for } \ \ -1<m<- m_{sp}(\beta)
 \nonumber \\
&& V'(m) >0 \ \ \ \ {\rm for } \ \ -m_{sp}(\beta)<m< 0
 \nonumber \\
&&  V'(m) <0 \ \ \ \ {\rm for } \ \ 0<m< m_{sp}(\beta)
 \nonumber \\
&& V'(m) >0 \ \ \ \ {\rm for } \ \ m_{sp}(\beta)<m< 1
\label{vprime}
\end{eqnarray}

The convergence towards the equilibrium distribution
of Eq. \ref{Psmall} for the continuous intensive magnetization $m \in [-1,+1]$
will be analyzed for the appropriate detailed-balance diffusion process in section \ref{sec_DBdiff}
and for the appropriate skew-detailed-balance run-and-tumble process in section \ref{sec_SkewDBrun}.


\section{  Detailed-Balance Markov chain for the magnetization $M \in [-N,..,+N]$ }

\label{sec_DBjump}

\subsection{ Single-spin-flip Markov chain satisfying Detailed-Balance in the space of the $2^N$ configurations}

The single-spin-flip master equation 
\begin{eqnarray}
 \partial_t {\cal P}_t( C )    =   \sum_{i=1 }^N \left[ {\cal W}(C,\sigma_i^x C ) {\cal P}_t(\sigma_i^x C) 
 -  {\cal W}(\sigma_i^x C ,C) {\cal P}_t(C) \right]
\label{markovchainDB}
\end{eqnarray}
has the following meaning.
From the configuration $C=\{S_1,..,S_i,.,S_N\}$, the possible elementary moves are towards
the $N$ configurations $(\sigma_i^x C)=\{S_1,..,-S_i,.,S_N\} $
obtained by the flip of a single spin $S_i$ towards $(-S_i)$,
and they occur with the rates ${\cal W}(\sigma_i^xC,C) $.

The simplest way to ensure the convergence of Eq. \ref{markovchainDB}
towards the thermal equilibrium of Eq. \ref{Pconfig} 
is to choose rates that satisfy Detailed-Balance, i.e.
where the current on each link vanishes in the steady state
\begin{eqnarray}
0= {\cal W}_{DB}(C,\sigma_i^x C ) {\cal P}_{eq}(\sigma_i^x C) -  {\cal W}_{DB}(\sigma_i^x,C) {\cal P}_{eq}(C) 
\label{markovchainDBc}
\end{eqnarray}
Following previous studies \cite{CurieW_slowingdown,lecomte_thermo,vivien_thesis},
we will consider the simplest rates satisfying Detailed-Balance 
\begin{eqnarray}
 {\cal W}_{DB}(\sigma_i^x C,C)  = e^{- \beta S_i \frac{M(C)}{N} } 
\label{DBchoice}
\end{eqnarray}
i.e. more explicitly

(i) if $S_i=-1$, the flip is towards $S_i=+1$ and the corresponding rate reads
\begin{eqnarray}
 {\cal W}_{DB}(\sigma_i^+ C,C)  = e^{ \beta J \frac{M(C)}{N} } 
\label{sigmap}
\end{eqnarray}

(ii) if $S_i=+1$, the flip is towards $S_i=-1$, and the corresponding rate reads
\begin{eqnarray}
 {\cal W}_{DB}(\sigma_i^- C,C)  = e^{ - \beta J \frac{M(C)}{N} } 
\label{sigmam}
\end{eqnarray}


\subsection{ Markov chain satisfying Detailed-Balance for the discrete extensive magnetization $M\in \{-N,..,N-2,N\}$ }

The dynamics in the configuration space of Eq. \ref{markovchainDB}
can be projected onto 
the one-dimensional dynamics for the $(N+1)$ possible values of the magnetization $M\in \{-N,..,N-2,N\}$ 
of Eq. \ref{even}
\begin{eqnarray}
  \partial_t P_{t}( M )  &&  =   \left[ W^+(M-2) P_t( M-2) - W^+(M) P_t(M) \right]
    + \left[ W^-(M+2) P_t( M+2) - W^-(M) P_t(M) \right]
\label{mchain}
\end{eqnarray}
with the  transition rates
\begin{eqnarray}
 W^+(M)  && = N_- e^{ \beta J \frac{M}{N} } = \left(\frac{N-M}{2}\right) \ e^{ \beta J \frac{M}{N} } 
 \nonumber \\
 W^-(M) && = N_+ e^{ \beta J \frac{M}{N} } = \left( \frac{N+M}{2} \right) \ e^{ - \beta J \frac{M}{N} } 
\label{Wm}
\end{eqnarray}
These rates satisfy the Detailed-Balance condition with respect to the equilibrium distribution of Eq. \ref{Pnpeq}
as it should for consistency
\begin{eqnarray}
\frac{ W^-(M+2) P_{eq}( M+2)}{  W^+(M) P_{eq}(M) }  =1
\label{mchainDB}
\end{eqnarray}
The magnetization dynamics of Eq. \ref{mchain} is of the form of Eq. \ref{master}
\begin{eqnarray}
  \partial_t P_{t}( M )    =  \sum_{M'} w(M,M')  P_t(M') 
\label{mchainmar}
\end{eqnarray}
with the Markov matrix
\begin{eqnarray}
w(M,M') = \left[ \delta_{M,M'+2} - \delta_{M,M'} \right] W^+(M')
+ \left[ \delta_{M,M'-2} - \delta_{M,M'} \right] W^-(M')
\label{markovwDB}
\end{eqnarray}
The Green function governing the typical convergence properties (see the reminder in Appendix \ref{sec_green}) 
is given in Appendix \ref{app_greenDBjump},
while here we apply the framework summarized in section \ref{sec_largedev}
to analyze the large deviations properties.


\subsection{ Large deviations at Level 2.5 for the empirical density $\rho(M)$ and the empirical flows $q(M \pm 2,M) $}

For the magnetization dynamics of Eq. \ref{mchain},
the large deviations at Level 2.5 (see subsection \ref{app_2.5})
involve the empirical magnetization density 
\begin{eqnarray}
 \rho(M) \equiv  \frac{1}{T} \int_0^T dt  \delta_{M(t),M}  
\label{empiMbis}
\end{eqnarray} 
and the empirical magnetization flows from $M$ towards $M \pm 2$
\begin{eqnarray}
q(M \pm 2,M) \equiv  \frac{1}{T} \sum_{t : M(t^+) \ne M(t^-)} \delta_{M(t^+),M \pm 2} \delta_{M(t^-),M} 
\label{jumpempiricaldensitym}
\end{eqnarray}

The joint probability distribution of the empirical density $\rho(.)$ and the empirical flows $q(.,.)$
follows the large deviation form of Eq \ref{level2.5}
\begin{eqnarray}
&& P^{Level \ 2.5}_{T}[ \rho(.) ; q(.,.) ]  \oppropto_{T \to +\infty} e^{- T I_{2.5}[ \rho(.) ; q(.,.) ] }
\nonumber \\
&& 
\delta \left( \sum_{ M=-N }^N   \rho(M)  - 1 \right) 
\prod_{ M=-N }^N \delta \left[ q(M - 2,M) + q(M + 2,M)
- q(M,M - 2)-q(M,M + 2)  \right]
\label{level2.5master}
\end{eqnarray}
with the rate function of Eq \ref{rate2.5}
\begin{eqnarray}
&& I_{2.5}[ \rho(.) ; q(.,.) ] =  \sum_{ M=-N }^{N-2}
\left[ q(M+2,M)  \ln \left( \frac{ q(M+2,M)  }{  W^+(M)  \rho( M) }  \right) 
 - q(M+2,M) + W^+(M)  \rho( M)  \right]
\nonumber \\
&&  +
 \sum_{ M=-N }^{N-2}
\left[ q(M,M+2)  \ln \left( \frac{ q(M,M+2)  }{  W^-(M+2)  \rho( M+2) }  \right) 
 - q(M,M+2) + W^-(M+2)  \rho( M+2)   \right]
\label{rate2.5master}
\end{eqnarray}
The stationarity constraints for flows in Eq. \ref{level2.5master}
means that the current on each link 
takes the same value $j$ for any magnetization $M$
\begin{eqnarray}
 q( M,M-2)-   q(M-2,M)
 = q(M+2,M)-  q( M,M+2) = j
 \label{jq}
\end{eqnarray}
and the boundary conditions at $M=\pm N$ with no flows imply that this current vanishes $j=0$.
As a consequence, it is convenient to replace the two flows on each link
by the new function $q_s$ of their middle-point
\begin{eqnarray}
q( M,M+2) = q(M+2,M) = q_s(M+1)  
 \label{qs}
\end{eqnarray}
in order to rewrite the level 2.5 of Eq. \ref{level2.5master}
as
\begin{eqnarray}
P_T^{Level \ 2.5} [ \rho(.) ; q_s(.)]
  \opsimeq_{T \to +\infty}  \delta \left( \sum_{ M=-N}^N   \rho(M)  - 1 \right) 
   e^{ - T I_{2.5}  [ \rho(.) ; q_s(.)] }
\label{level2.5qd}
\end{eqnarray}
with the rate function
\begin{eqnarray}
&& I_{2.5}[ \rho(.) ; q_s(.) ] =  \sum_{ M=-N }^{N-2}
\left[ q_s(M+1)   \ln \left( \frac{ q_s(M+1)   }{  W^+(M)  \rho( M) }  \right) 
 - q_s(M+1)  + W^+(M)  \rho( M)  \right]
\label{rate2.5qs} \\
&&  +
 \sum_{ M=-N }^{N-2}
\left[ q_s(M+1)   \ln \left( \frac{ q_s(M+1)  }{  W^-(M+2)  \rho( M+2) }  \right) 
 - q_s(M+1)  + W^-(M+2)  \rho( M+2)   \right]
 \nonumber \\
 &&
 =  \sum_{ M=-N }^{N-2}
\left[ q_s(M+1)   \ln \left( \frac{ q_s^2(M+1)   }{  W^+(M)  \rho( M)W^-(M+2)  \rho( M+2) }  \right) 
 - 2 q_s(M+1)  + W^+(M)  \rho( M) + W^-(M+2)  \rho( M+2) \right]
\nonumber
\end{eqnarray}


\subsection{ Large deviations at Level 2 for the empirical density $\rho(M)$ alone }

For a given empirical density $\rho(M)$,
the optimization of the rate function of Eq. \ref{rate2.5qs}
with respect to the link flow $q_s(M+1)$
\begin{eqnarray}
&&  0 = \frac{ \partial   I_{2.5}[ \rho(.) ; q_s(.) ]  }{\partial q_s(M+1) }
=    \ln \left( \frac{ q_s^2(M+1)   }{  W^+(M)  \rho( M)W^-(M+2)  \rho( M+2) }  \right) 
 \label{rate2.5qsderi}
\end{eqnarray}
yields the explicit optimal value
\begin{eqnarray}
q^{opt}_s(M+1)  = \sqrt{  W^+(M)  \rho( M)W^-(M+2)  \rho( M+2) }
 \label{rate2.5deri}
\end{eqnarray}
as a function of the given empirical density $\rho(M)$.
So
the probability distribution of the empirical density $\rho(.)$ alone
follows the large deviation form
\begin{eqnarray}
P_T^{Level \ 2} [ \rho(.) ]
  \opsimeq_{T \to +\infty}  \delta \left( \sum_{ M=-N }^N   \rho(M)  - 1 \right) 
   e^{ - T I_{2}  [ \rho(.)] }
\label{level2master}
\end{eqnarray}
where the rate function at Level 2 obtained 
from the optimal value of the rate function at the Level 2.5 of Eq. \ref{rate2.5qs} is explicit
\begin{eqnarray}
&& I_{2}  [ \rho(.)]  = I_{2.5}[ \rho(.) ; q^{opt}_s(.) ] 
\label{rate2master}
 \\
&&  =  \sum_{ M=-N }^{N-2}
\left[  W^+(M)  \rho( M) + W^-(M+2)  \rho( M+2) 
-2  \sqrt{ W^+(M)  \rho( M)W^-(M+2)  \rho( M+2) }
\right]
\nonumber \\
&& =  \sum_{ M=-N }^{N-2}
\left[   \sqrt{ W^+(M)  \rho( M) } -  \sqrt{ W^-(M+2)  \rho( M+2) }
\right]^2
\nonumber
\end{eqnarray}
In the last expression, the detailed-balance condition of Eq. \ref{mchainDB}
can be used to replace $W^-(M+2)= \frac{W^+(M) P_{eq}(M)}{P_{eq}( M+2)}$ to obtain
the final result
\begin{eqnarray}
 I_{2}  [ \rho(.)]  =  \sum_{ M=-N }^{N-2}  W^+(M) P_{eq}( M)  
\left[   \sqrt{ \frac{  \rho( M) }{ P_{eq}( M)} } -  \sqrt{ \frac{  \rho( M+2)}{P_{eq}( M+2)} }   \right]^2
\end{eqnarray}
where the vanishing for $\rho(M)=P_{eq}(M)$ is obvious.
The closed form of this rate function at Level 2 is a specific example of the standard theory of
Donsker and Varadhan \cite{DonskerV} for Markov jump processes satisfying Detailed-Balance.
Here the Level 2 has been derived from the contraction of the Level 2.5 in order to have a unified framework 
with the other sections concerning Markov processes breaking Detailed-Balance.


\section{ Skew-Detailed-Balance Markov chain for the magnetization $M \in [-N,..,+N]$}

\label{sec_SkewDBjump}


\subsection{ Single-spin-flip Markov chain satisfying Skew-Detailed-Balance in the space of configurations}

For the present Curie-Weiss model, the skew-detailed-balance algorithm
introduced by Konstantin Turitsyn, Michael Chertkov and Marija Vucelja in \cite{TCV}
(see the review \cite{vuceljaa} and references therein)
 can be summarized as follows.
Besides the configuration $C=\{S_1,..,S_i,.,S_N\}$ of the $N$ spins, 
one introduces the supplementary variable $\sigma=\pm$
that will restrict the next possible moves as follows :

(i) when $\sigma=+1$, only the single-spin-flip towards configurations $\sigma_i^+C$ of higher magnetization are possible.

(ii) when $\sigma=-1$, only the single-spin-flip towards configurations $\sigma_i^-C$ of lower magnetization are possible.

In addition, the supplementary variable can flip $\sigma \to - \sigma$ with some switching rates $\Gamma^{\sigma}(C)$
without changing the configuration $C$.
As a consequence, the master equations for the probability $P_t^{\sigma}(C)$ to be in configuration $C$ 
and in the copy $\sigma$ at time $t$ are of the form
\begin{eqnarray}
 \partial_t {\cal P}^+_t( C ) &&   =  
  \sum_{i : S_i=+1 }  {\cal W}^+(C,\sigma_i^- C ) {\cal P}^+_t(\sigma_i^- C)
  -  \sum_{i : S_i=-1 } {\cal W}^+(\sigma_i^+ C,C) {\cal P}^+_t(C) 
  -   \Gamma^+(C) {\cal P}^+_t(C) +  \Gamma^-(C) {\cal P}^-_t(C)
 \nonumber \\
  \partial_t {\cal P}^-_t( C ) &&   = 
    \sum_{i : S_i=-1 }  {\cal W}^-(C,\sigma_i^+ C ) {\cal P}^-_t(\sigma_i^+ C)
  -  \sum_{i : S_i=+1 } {\cal W}^-(\sigma_i^- C,C) {\cal P}^-_t(C) 
  +   \Gamma^+(C) {\cal P}^+_t(C) -  \Gamma^-(C) {\cal P}^-_t(C)
\label{lifted}
\end{eqnarray}
where the rates have to be chosen in order to ensure the convergence towards the steady state
 related to the equilibrium distribution $ {\cal P}_{eq}(C)$ of \ref{Pconfig} 
 \begin{eqnarray}
{\cal P}_*^{+}(C)  ={\cal P}_*^{-}(C)  = \frac{{\cal P}_{eq}(C)}{2} 
\label{steadyLifted}
\end{eqnarray}
So the rates have to satisfy
\begin{eqnarray}
0 &&   =  
  \sum_{i : S_i=+1 }  {\cal W}^+(C,\sigma_i^- C ) {\cal P}_{eq}(\sigma_i^- C)
  -  \sum_{i : S_i=-1 } {\cal W}^+(\sigma_i^+ C,C) {\cal P}_{eq}(C) 
  -   \Gamma^+(C) {\cal P}_{eq}(C) +  \Gamma^-(C) {\cal P}_{eq}(C)
 \nonumber \\
 0 &&   = 
   \sum_{i : S_i=-1 }  {\cal W}^-(C,\sigma_i^+ C ) {\cal P}_{eq}(\sigma_i^+ C)
  -  \sum_{i : S_i=+1 } {\cal W}^-(\sigma_i^- C,C) {\cal P}_{eq}(C) 
+   \Gamma^+(C) {\cal P}_{eq}(C) -  \Gamma^-(C) {\cal P}_{eq}(C)
\label{liftedeq}
\end{eqnarray}
and the choices proposed by Konstantin Turitsyn, Michael Chertkov and Marija Vucelja in \cite{TCV} (see the review \cite{vuceljaa} and references therein)
can be explained as follows.

(a) The sum of these two equations allows to eliminate the switching rates $ \Gamma^{\pm}(C)$
\begin{eqnarray}
0  &&  = 
  \sum_{i : S_i=+1 } \left[ {\cal W}^+(C,\sigma_i^- C ) {\cal P}_{eq}(\sigma_i^- C)
  -  {\cal W}^-(\sigma_i^- C,C) {\cal P}_{eq}(C)  \right]
\nonumber \\
&& + 
   \sum_{i : S_i=-1 } \left[ {\cal W}^-(C,\sigma_i^+ C ) {\cal P}_{eq}(\sigma_i^+ C)
 -   {\cal W}^+(\sigma_i^+ C,C) {\cal P}_{eq}(C) \right]
\label{liftedeqsum}
\end{eqnarray}
 The simplest way to satisfy these equations is to impose the
Skew-Detailed-Balance \cite{TCV} for the spin-flip rates ${\cal W}^{\pm} (.,.)$
\begin{eqnarray}
0 =  {\cal W}^+_{SDB}(C,\sigma_i^- C ) {\cal P}_{eq}(\sigma_i^- C)
  -  {\cal W}^-_{SDB}(\sigma_i^- C,C) {\cal P}_{eq}(C)
\label{skewDB}
\end{eqnarray}
The physical meaning is that the total current after summing over the two copies vanishes.
As a consequence, one can choose the rates as in Eqs \ref{DBchoice} with Eqs \ref{sigmap}
concerning the detailed-balance dynamics
and \ref{sigmam}
\begin{eqnarray}
 {\cal W}^+(\sigma_i^+ C,C) && =e^{ \beta J \frac{M(C)}{N} } 
 \nonumber \\
  {\cal W}^-(\sigma_i^- C,C) && =  e^{ - \beta J \frac{M(C)}{N} } 
\label{skewDBchoice}
\end{eqnarray}

(b) Rewriting the second equation of Eqs \ref{liftedeq}
using Eq. \ref{skewDB} for the first term,
one obtains that ${\cal P}_{eq}(C) $ appears in factor of all the terms,
so that what remains is the following constraint for the difference between the two 
positive switching rates $ \Gamma^{\pm}(C) \geq 0$
\begin{eqnarray}
    \Gamma^+(C)  -  \Gamma^-(C) && = 
      \sum_{i : S_i=+1 } {\cal W}^-(\sigma_i^- C,C) 
  -  \sum_{i : S_i=-1 } {\cal W}^+(\sigma_i^+ C,C) 
\nonumber \\
 && =  \left( \frac{N+M(C)}{2} \right) \ e^{ - \beta J \frac{M(C)}{N} }  - 
  \left(\frac{N-M(C)}{2}\right) \ e^{ \beta J \frac{M(C)}{N} } 
\label{diffgamma}
\end{eqnarray}
The choice of \cite{TCV} is as follows : for each configuration $C$, one imposes that one of the two 
switching rates $ \Gamma^{\pm}(C) $ vanishes, so that it is the sign and the amplitude of the right hand site of 
Eq. \ref{diffgamma} that determines which switching rate does not vanish and what is its value 
\begin{eqnarray}
  \Gamma^+(C)    && =  {\rm max} \left[ 0, \left( \frac{N+M(C)}{2} \right) \ e^{ - \beta J \frac{M(C)}{N} }  - 
  \left(\frac{N-M(C)}{2}\right) \ e^{ \beta J \frac{M(C)}{N} }  \right]
  \nonumber \\  
    \Gamma^-(C)   &&  =  {\rm max} \left[ 0, - \left( \frac{N+M(C)}{2} \right) \ e^{ - \beta J \frac{M(C)}{N} }  +
  \left(\frac{N-M(C)}{2}\right) \ e^{ \beta J \frac{M(C)}{N} }    \right]
\label{choicegamma}
\end{eqnarray}


\subsection{ Corresponding Skew-Detailed-Balance Lifted-Markov-Chain for the magnetization $M \in [-N,..,+N]$ }

The lifted dynamics of Eq. \ref{lifted} can be projected onto the following dynamics for the magnetization $M$
\begin{eqnarray}
  \partial_t P_{t}^+( M )  &&  =    W^+(M-2) P_t^+( M-2)
 - W^+(M) P_t^+(M) - \Gamma^+(M)P_{t}^+( M )+ \Gamma^-(M)P_{t}^-( M )
 \nonumber \\
   \partial_t P_{t}^-( M )  &&  =    W^-(M+2) P_t^-( M+2)
 - W^-(M) P_t^-(M) + \Gamma^+(M)P_{t}^+( M ) - \Gamma^-(M)P_{t}^-( M )
\label{liftedproj}
\end{eqnarray}
with the following rates.

\subsubsection{ Skew-Detailed-Balance for the rates $W^{\pm}(M)$ governing the changes in magnetization in each copy  }

 The rates $W^{\pm}(.)$ governing the changes in magnetization are given by the same expression as in Eqs \ref{Wm}
\begin{eqnarray}
 W^+(M)  && = N_- e^{ \beta J \frac{M}{N} } = \left(\frac{N-M}{2}\right) \ e^{ \beta J \frac{M}{N} } 
 \nonumber \\
 W^-(M) && = N_+ e^{ \beta J \frac{M}{N} } = \left( \frac{N+M}{2} \right) \ e^{ - \beta J \frac{M}{N} } 
\label{Wmbis}
\end{eqnarray}
So they satisfy the corresponding Skew-Detailed-Balance condition (Eq. \ref{skewDB})
\begin{eqnarray}
 0=  W^-(M+2) P_{eq}( M+2) - W^+(M) P_{eq}(M)
\label{skewDBproj}
\end{eqnarray}
that ensures the convergence towards the steady-state similar to Eq. \ref{steadyLifted}
 \begin{eqnarray}
P_*^{+}(M)  =P_*^{-}(M)  = \frac{P_{eq}(M)}{2} 
\label{steadyLiftedprof}
\end{eqnarray}

\subsubsection{Switching rates $\Gamma^{\pm}(M)$ between the two copies at magnetization $M$ }

 The switching rates $\Gamma^{\pm}(.)$ obtained from Eq. \ref{choicegamma} read
\begin{eqnarray}
  \Gamma^+(M)    && =   {\rm max} \left[ 0, W^-(M)-W^+(M)  \right]
  \nonumber \\  
    \Gamma^-(M)   &&  = {\rm max} \left[ 0, W^+(M)-W^-(M) \right]
\label{gamman}
\end{eqnarray}
In particular, they satisfy the analog of Eq. \ref{diffgamma}
\begin{eqnarray}
    \Gamma^+(M)  -  \Gamma^-(M) && = W^-(M)-W^+(M) 
\label{diffgammam}
\end{eqnarray}
or equivalently
\begin{eqnarray}
    \Gamma^+(M) + W^+(M)  =  \Gamma^-(M) + W^-(M) 
\label{diffgammaout}
\end{eqnarray}
with the following physical meaning : the total rate $( \Gamma^+(M) + W^+(M))$ out of the state $(M,\sigma=+)$
should be equal to the total rate $( \Gamma^-(M) + W^-(M))$ out of the state $(M,\sigma=-)$.

In order to find the biggest rate between $W^{+}(M)$ and $W^-(M)$ of Eq. \ref{Wmbis},
it is useful to rewrite their ratio in terms of the derivative $V'(m)$ of Eq. \ref{vmderi}
\begin{eqnarray}
\frac{ W^-(M)  }{ W^+(M)} && 
= \frac{ \left( \frac{N+M}{2} \right) \ e^{ - \beta J \frac{M}{N} } }{ \left(\frac{N-M}{2}\right) \ e^{ \beta J \frac{M}{N} } }
= \frac{1+\frac{M}{N} } {1- \frac{M}{N} } e^{- 2 \beta J \frac{M}{N} } = e^{2 V' \left(\frac{M}{N} \right) }
\label{Wmratio}
\end{eqnarray}
As a consequence, the sign of the derivative $V' \left(\frac{M}{N} \right) $ 
determines which switching rate vanishes,
and one can rewrite Eqs \ref{gamman}
 in terms of the Heaviside theta function $\theta(.)$ of the derivative $V'(m)$
\begin{eqnarray}
  \Gamma^+(M)    && = \theta\left(V' \left(\frac{M}{N} \right) \right)   \left[ W^-(M)-W^+(M)  \right]
  \nonumber \\  
    \Gamma^-(M)   &&  = \theta\left(- V' \left(\frac{M}{N} \right) \right)   \left[ W^+(M)-W^-(M) \right]
\label{gammathetaV}
\end{eqnarray}
Since the regions where the derivative $V'(m)$ is positive or negative
are different in the high-temperature phase (see Eq. \ref{vprimehigh})
and in the low-temperature phase (see Eq. \ref{vprime}),
it is useful to write more explicitly the two cases :

(i) in the high-temperature phase $\beta J < \beta_c J=1$ or exactly at criticality,
the switching rates of Eq. \ref{gammathetaV} read (see Eq. \ref{vprimehigh})
\begin{eqnarray}
&&   \Gamma^+(M) = 0  \ \ \ \ \ { \rm and } \ \ \ \  \ \ \  \Gamma^-(M)  =W^+(M)-W^-(M) 
\ \  \ \ \  {\rm for } \ \ M<0
  \nonumber \\  
&&  \Gamma^+(M) = W^-(M)-W^+(M)  \ \ \ \ \ \ { \rm and } \ \ \ \ \ \ \Gamma^-(M) =0
\ \  \ \ \ {\rm for } \ \ M>0
\label{gammahigh}
\end{eqnarray}

(ii) in the low-temperature phase $\beta J > \beta_c J=1$,
the switching rates of Eq. \ref{gammathetaV} read (see Eq. \ref{vprime})
\begin{eqnarray}
&&   \Gamma^+(M) = 0  \ \ \ \ \ { \rm and } \ \ \ \  \ \ \  \Gamma^-(M)  =W^+(M)-W^-(M) 
\ \  \ \ \ {\rm for } \ \ M< -N m_{sp}(\beta) \ \ \  { \rm and } \ \ 0<M<N m_{sp}(\beta)
  \nonumber \\  
&&   \Gamma^+(M) = W^-(M)-W^+(M)  \ \ \ \ \ \ { \rm and } \ \ \ \ \ \ \Gamma^-(M) =0
\ \  \ \ \  {\rm for } \   -N m_{sp}(\beta)<M<0 \ \ { \rm and } \ \ N m_{sp}(\beta) <M
\label{gammalow}
\end{eqnarray}


\subsubsection{ Summary : Markov matrix for the Directed Markov jump process on the ladder }

The dynamics of Eq. \ref{liftedproj} can be rewritten as Eq. \ref{master}
where the state $x=(M,\sigma)$ contains both the Magnetization $M$ and the supplementary variable $\sigma=\pm 1$
\begin{eqnarray}
  \partial_t P_{t}^{\sigma}( M )  &&  =   \sum_{\sigma'=\pm 1} \sum_{M'} w^{\sigma,\sigma'}_{M,M'} P_{t}^{\sigma'}( M' )
\label{liftedmasterw}
\end{eqnarray}
with the Markov matrix 
\begin{eqnarray}
 w^{\sigma,\sigma'}_{M,M'} && = 
 \delta^{\sigma,+} \delta^{\sigma',+}   \left[ \delta_{M,M'+2} -\delta_{M,M'} \right] W^+(M') 
 +  \delta^{\sigma,-} \delta^{\sigma',-}   \left[ \delta_{M,M'-2} - \delta_{M,M'} \right] W^-(M') 
\nonumber \\
&&  + \left[ \delta^{\sigma,+}   - \delta^{\sigma,-}  \right] \delta^{\sigma',-} \delta_{M,M'} \Gamma^-(M')
 + \left[ \delta^{\sigma,-}    - \delta^{\sigma,+}  \right] \delta^{\sigma',+}\delta_{M,M'} \Gamma^+(M')
\label{wblock}
\end{eqnarray}
The enlarged configuration space $(M,\sigma=\pm 1)$ is thus a ladder, 
and the flows are possible only in one direction on each link.
The Green function governing the typical convergence properties (see the reminder in Appendix \ref{sec_green}) 
is computed in Appendix \ref{app_greenSDBjump},
while here we apply the framework summarized in section \ref{sec_largedev}
to analyze the large deviations properties.


\subsection{ Explicit large deviations at Level 2.5 for the empirical densities $\rho^{\pm}(M)$ and the empirical flows  }

For the present dynamics, the large deviations at Level 2.5 of Eq. \ref{level2.5} 
involve the following empirical observables with their constraints :

(i) The two empirical densities of the magnetization $M$ in the two copies $\sigma=\pm 1$
\begin{eqnarray}
 \rho^{ \sigma}( M)  \equiv \frac{1}{T} \int_0^T dt  \delta_{\sigma(t),\sigma}  \ \delta_{  M(t),  M} 
 \label{rho1def}
\end{eqnarray}
  satisfy the global normalization
\begin{eqnarray}
  \sum_{ M=-N }^N  \left[ \rho^{ +}( M) + \rho^{ -}( M)\right]=1
\label{rho1norma}
\end{eqnarray}

(ii) The magnetization current in the copy $\sigma=+$
between two neighboring magnetizations $M \to M + 2$
\begin{eqnarray}
 j^+(M+1)  \equiv  \frac{1}{T} \sum_{t : M(t^+) \ne M(t^-)} \delta_{\sigma(t),+}
 \delta_{M(t^+),M+2} \delta_{M(t^-),M} 
\label{jp}
\end{eqnarray}
and the magnetization current in the copy $\sigma=-$
between two neighboring magnetizations $M \to M - 2$
\begin{eqnarray}
 j^-(M-1)  \equiv  \frac{1}{T} \sum_{t : M(t^+) \ne M(t^-)} \delta_{\sigma(t),-}
 \delta_{M(t^+),M-2} \delta_{M(t^-),M} 
\label{jpm}
\end{eqnarray}
are labelled by their middle-point $(M \pm 1)$ to simplify the notations.

(iii) At magnetization $M$, the switching events between the two copies
are described by the empirical switching flows  
\begin{eqnarray}
Q^+( M)  \equiv  \frac{1}{T} \sum_{t  \in [0,T]: \sigma(t^+)=- \ne \sigma(t^-)=+ }  \  \delta_{M(t),  M} 
\nonumber \\
Q^-( M)  \equiv  \frac{1}{T} \sum_{t  \in [0,T]: \sigma(t^+) =+\ne \sigma(t^-)=- }   \delta_{M(t),  M}
\label{jumpflows}
\end{eqnarray}

(iv) The stationarity conditions mean that the total flows into the state 
$(M,\sigma)$ should be balanced by the total flows out of the state $(M,\sigma)$.
These conditions read for magnetization $M$ and the two possible states $\sigma=\pm 1$
\begin{eqnarray}
  0  &&  =    j^+(M-1)  - j^+(M+1)  - Q^+(M)+Q^-(M)
 \nonumber \\
0  &&  =    j^-(M+1)  - j^-(M-1)  + Q^+(M)-Q^-(M)
\label{statioprojproj}
\end{eqnarray}
The sum of these two equations allows to eliminate the two switching flows $Q^{\pm}(M)$
and yields that the total magnetization current takes the same $j$ on each link
\begin{eqnarray}
  j^+(M-1) -  j^-(M-1)  =  j^+(M+1) -   j^-(M+1) = j
 \label{jconserv}
\end{eqnarray}
and thus vanishes $j=0$ as a consequence of the boundary conditions at $M=\pm N$ without flows.
As a consequence, it will be possible to eliminate all the negative 
currents $j^-(.)$ in terms of the positive current $j^+(.)$
\begin{eqnarray}
 j^-(M+1) =  j^+(M+1) 
 \label{jmjp}
\end{eqnarray}
while the remaining stationarity condition of Eq. \ref{statioprojproj} involving the switching flows $Q^{\pm}(M)$ read
\begin{eqnarray}
    Q^-(M) -   Q^+(M)   =  j^+(M+1) -     j^+(M-1) 
 \label{jqst}
\end{eqnarray}
For each magnetization $M$, one of the two switching flows $Q^{\pm}(M) $ 
is not possible as a consequence of the choice of 
the two switching rates $\Gamma^{\pm}(M)$
Eq. \ref{gammathetaV}.
So the stationarity condition of Eq. \ref{jqst}
can be rewritten more explicitly as
\begin{eqnarray}
        Q^+(M)   =  j^+(M-1) - j^+(M+1)      \ \ \ {\rm for } \ \ V' \left(\frac{M}{N} \right) >0
    \nonumber \\
      Q^-(M)    =  j^+(M+1) -     j^+(M-1) \ \ \ {\rm for } \ \ V' \left(\frac{M}{N} \right) <0 
 \label{elimQjp}
\end{eqnarray}

The application of the Level 2.5 of Eq. \ref{level2.5} 
yields that the joint distribution of the two empirical density $ \rho^{\pm}(.) $,
of the two empirical currents $ j^{\pm}(.) $
and of the two switching flows $ Q^{\pm}(.)$
reads
\begin{eqnarray}
 && P_T^{Level \ 2.5}[ \rho^{\pm}(.), j^{\pm}(.) , Q^{\pm}(.) ]   \opsimeq_{T \to +\infty}   
 e^{- \displaystyle T  I_{2.5}[\rho_{\pm}(.), j^{\pm}(.),  Q^{\pm}(.) ]   }
     \nonumber \\
&&
  \delta \left( \sum_{ M=-N }^N \left[ \rho^{ +}( M) + \rho^{ -}( M)\right] -1 \right)
\left[ \prod_{ M=-N }^{N-2} \delta \left(   j^-(M+1) -  j^+(M+1)      \right) \right]
    \nonumber \\
&&\left[ \prod_{ M : V' \left(\frac{M}{N} \right)>0 } \delta \left(      j^+(M+1) -     j^+(M-1) + Q^+(M)      \right)  \right] 
\left[ \prod_{ M: V' \left(\frac{M}{N} \right)<0 } \delta \left(        j^+(M+1) -     j^+(M-1) -Q^-(M)      \right)  \right] 
\label{ld2.5rhoq}
\end{eqnarray}
with the explicit rate function at Level 2.5
\begin{eqnarray}
&& I_{2.5}[ \rho_{\pm}(.), j^{\pm}(.),  Q^{\pm}(.) ]   = 
  \sum_{ M=-N }^{N-2}
\left[  j^+(M+1)   \ln \left( \frac{  j^+(M+1)   }{  W^+(M) \rho^+(M)  }  \right) 
 -    j^+(M+1)   +   W^+(M) \rho^+(M)     \right]  
\nonumber \\
&& +   \sum_{ M=-N}^{N-2}
\left[  j^-(M+1)   \ln \left( \frac{  j^-(M+1)   }{  W^-(M+2) \rho^-(M+2)  }  \right) 
 -    j^-(M+1)   +   W^-(M+2) \rho^-(M+2)    \right]  
\nonumber \\
&& 
 +  \sum_{ M: V' \left(\frac{M}{N} \right)>0 }
\left[  Q^{+}( M)    \ln \left( \frac{  Q^{+}( M)   }{ \left[ W^-(M)-W^+(M)  \right] \rho^+(M)  }  \right) 
 -   Q^{+}( M)   +   \left[ W^-(M)-W^+(M)  \right] \rho^+(M)     \right]  
\nonumber \\
&&  +  \sum_{ M: V' \left(\frac{M}{N} \right)<0 }
\left[  Q^{-}( M)    \ln \left( \frac{  Q^{-}( M)   }{ \left[ W^+(M)-W^-(M) \right] \rho^-(M)  }  \right) 
 -   Q^{-}( M)   +  \left[ W^+(M)-W^-(M) \right] \rho^-(M)     \right]  
\label{rate2.5rhoq}
\end{eqnarray}
Using Eqs \ref{jmjp} and \ref{elimQjp} that solve the stationarity conditions,
 one obtains that Eq. \ref{ld2.5rhoq}
reduces to the joint probability distribution of three variables only, namely
the two empirical density $ \rho^{\pm}(.) $,
and the positive currents $ j^{+}(.) $
\begin{eqnarray}
  P_T^{Level \ 2.5}[ \rho_{\pm}(.), j^{+}(.)  ]   \opsimeq_{T \to +\infty}   
  \delta \left( \sum_{ M=-N }^N \left[ \rho^{ +}( M) + \rho^{ -}( M)\right] -1 \right)
  e^{- \displaystyle T  I_{2.5}[\rho_{\pm}(.), j^{+}(.) ]   }
\label{ld2.5rhoqjp}
\end{eqnarray}
with the rate function
\begin{footnotesize}
\begin{eqnarray}
&& I_{2.5}[ \rho_{\pm}(.), j^{+}(.) ]   = 
 \label{rate2.5rhoqjp}
 \\
&&    \sum_{ M=-N }^{N-2}
\left[  j^+(M+1)   \ln \left( \frac{  [j^+(M+1)]^2   }{ W^+(M) \rho^+(M) W^-(M+2) \rho^-(M+2)  }  \right) 
 -   2 j^+(M+1) +   W^+(M) \rho^+(M)  +   W^-(M+2) \rho^-(M+2)    \right]  
\nonumber \\
&& 
 +  \sum_{ M: V' \left(\frac{M}{N} \right) >0 }
\left[   [j^+(M-1) - j^+(M+1)       ]    \ln \left( \frac{   [j^+(M-1)- j^+(M+1)  ]   }{  \left[ W^-(M)-W^+(M)  \right] \rho^+(M)  }  \right) 
 + j^+(M+1) -     j^+(M-1) ]   +   \left[ W^-(M)-W^+(M)  \right] \rho^+(M)     \right]  
\nonumber \\
&&  +  \sum_{ M: V' \left(\frac{M}{N} \right)<0 }
\left[  [j^+(M+1) -     j^+(M-1) ]    \ln \left( \frac{  [j^+(M+1) -     j^+(M-1) ]   }{  \left[ W^+(M)-W^-(M) \right] \rho^-(M)  }  \right) 
 -   j^+(M+1) +     j^+(M-1) ]   +   \left[ W^+(M)-W^-(M) \right] \rho^-(M)     \right]  
\nonumber
\end{eqnarray}
\end{footnotesize}
So here the large deviations at level 2.5 are more complicated than for the detailed-balance case of the
previous section where it was easy to obtain the Level 2 via explicit contraction.
However, the large deviations properties will be simpler for the 
corresponding Skew-Detailed-Balance run-and-tumble process for the intensive continuous magnetization $m$
as described in section \ref{sec_SkewDBrun}.


\section{ Effective Detailed-Balance diffusion dynamics for $m \in ]-1,1[$  }

\label{sec_DBdiff}


\subsection{ Detailed-Balance diffusion dynamics for the intensive magnetization $m \in ]-1,1[$ for large $N$ }

If one wishes to replaces the Detailed-Balance Markov chain of Eq. \ref{mchain} for the discrete magnetization $M$
by the corresponding diffusion process for continuous $M$, one obtains the Fokker-Planck dynamics
\begin{eqnarray}
  \partial_t P_{t}( M )  &&  = 2 N \partial_M  \left[ \partial_M P_t(M) + V' \left( \frac{M}{N} \right) P_t(M) \right]
\label{diffmbig}
\end{eqnarray}
involving the diffusion coefficient $(2N)$ and the derivative $V'(m)$ discussed in Eq. \ref{vmderi}.
For the probability $p_t(m)$ of the intensive magnetization $m=\frac{M}{N}$, 
Eq. \ref{diffmbig} translates into the Fokker-Planck dynamics
\begin{eqnarray}
  \partial_t p_{t}( m )  &&  = \frac{2}{N} \partial_m  \left[ \partial_m p_t(m) + N V' (m) p_t(m)\right]
\label{diffmsmall}
\end{eqnarray}
that involves the diffusion coefficient $D = \left( \frac{2}{N} \right) $ and that converges towards the 
normalized equilibrium distribution (Eq. \ref{Psmall})
\begin{eqnarray}
  p_{eq}(m) = \frac{ e^{-  N  V(m) } }
  { \int_{-1}^{+1} dx e^{-  N  V(x) } }
\label{Peqm}
\end{eqnarray}
Here, the detailed-balance property means that the current vanishes in the steady state $p_{eq}(m)$
 \begin{eqnarray}
  j_{eq}(m)    = - \frac{2}{N}   \left[ \partial_m p_{eq}(m) + N V' (m) p_{eq}(m)\right] =0
\label{jeqzero}
\end{eqnarray}

In conclusion, Eq. \ref{diffmsmall} corresponds to the Fokker-Planck dynamics of Eq. \ref{fokkerplanck}
in dimension $d=1$ 
\begin{eqnarray}
  \partial_t p_{t}( m )  &&  =  \partial_m  \left[ D \partial_m p_t(m) + 2 V' (m) p_t(m)\right]
\label{fpdf1d}
\end{eqnarray}
where the diffusion coefficient $D$ is independent of $m$ (but depends on $N$)
 \begin{eqnarray}
D = \frac{2}{N}
\label{dcoef}
\end{eqnarray}
while the force $F(m)$ involves the derivative $V'(m)$ of the effective potential $V(m)$ of Eq. \ref{vm}
 \begin{eqnarray}
F(m) = - 2 V'(m) = - D N V'(m)
\label{forcem}
\end{eqnarray}
The Green function governing the typical convergence properties (see the reminder in Appendix \ref{sec_green}) 
is given in Appendix \ref{app_greenDBdiff},
while here we apply the framework of subsection \ref{subsec_2.5diff}
to analyze the large deviations properties.


\subsection{ Large deviations at Level 2 for the empirical density $\rho(m)$ }

For the present model, the large deviations at level 2.5 of Eq. \ref{ld2.5diff}
involve the stationarity constraint for the empirical magnetization current $j(m)$
\begin{eqnarray}
\frac{dj(m)}{dm} =0
\label{derijzero}
\end{eqnarray}
So the empirical current $j(m)$ cannot depend on $m$ 
and it thus vanishes $j(m)=0$ as a consequence of the boundaries without flows.
As a consequence, the large deviations at level 2.5 of Eq. \ref{ld2.5diff}
directly reduce to the large deviations at Level 2 
for the probability
of the empirical density $\rho(m) $ alone
\begin{eqnarray}
 P_T[ \rho(.)]   \opsimeq_{T \to +\infty}  \delta \left(\int_{-1}^{+1} dm \  \rho(m) -1  \right)
e^{- \displaystyle T I_{2} [ \rho(.)] }
\label{ld2.5diffm}
\end{eqnarray}
where the rate function at level 2 reads using the force $F(m) =  - D N V'(m) $ of Eq. \ref{forcem}
and the diffusion coefficient $D = \frac{2}{N}$ of Eq. \ref{dcoef}
\begin{eqnarray}
I_{2} [ \rho(.)]  && =
\int_{-1}^{+1} \frac{dm}{ 4 D  \rho(m) } \left[  - \rho(m)  F(m)+D   \rho'(m) \right]^2
\nonumber \\
&&= \frac{1}{2N}  \int_{-1}^{+1} dm    \rho(m)  \left[     N V'(m)+  \frac{ \rho'(m) }{\rho(m) } \right]^2
 =\frac{N}{2}  \int_{-1}^{+1} dm    \rho(m)  \left[      V'(m)+ \frac{1}{N} \left( \frac{ \rho'(m) }{ \rho(m) } \right) \right]^2
\label{rate2diff}
\end{eqnarray}
In order to compare more directly with the equilibrium distribution of Eq. \ref{Peqm},
it is convenient to parametrize the normalized empirical density $\rho(.)$ via the corresponding empirical potential $U(.)$
\begin{eqnarray}
  \rho(m) =   \frac{ e^{-  N  U(m) } }  { \int_{-1}^{+1} dx e^{-  N  U(x) } }
\label{rhomparaU}
\end{eqnarray}
so that the rate function of Eq. \ref{rate2diff} translates into
\begin{eqnarray}
I^{DB}_{2} \left[ \rho(.)=\frac{ e^{-  N  U(.) } }  { \int_{-1}^{+1} dx e^{-  N  U(x) } } \right]  
 =   \frac{ N  \int_{-1}^{+1} dm  e^{-  N  U(m) }  \left[  U'(m)-    V'(m) \right]^2}  { 2 \int_{-1}^{+1} dx e^{-  N  U(x) } } 
\label{rate2diffUpara}
\end{eqnarray}
In particular, the cost of a small perturbation $\epsilon(m)=U(m)-V(m)$ of the empirical potential $U(m)$ around the equilibrium potential $V(m)$ reads at lowest order
\begin{eqnarray}
 I_{2}^{DB} \left[ \rho(.)=\frac{ e^{-  N  \left[ V(.) + \epsilon(.) \right] } }  { \int_{-1}^{+1} dx e^{-  N  \left[ V(x) + \epsilon(x) \right] } } \right] 
 \opsimeq_{\epsilon(.) \to 0}
\frac{ N  \int_{-1}^{+1}d  m \ e^{-  N  V(m) } \ 
   [\epsilon'(m) ]^2  }
 { 2 \int_{-1}^{+1} dx e^{-  N  V(x) } } 
\label{rate2diffUparaperturb}
\end{eqnarray}


\section{ Effective Skew-Detailed-Balance run-and-tumble dynamics for $m \in [-1,1]$  }

\label{sec_SkewDBrun}


\subsection{ Skew-Detailed-Balance run-and-tumble dynamics for the intensive magnetization $m \in ]-1,1[$ for large $N$ }

If one wishes to replace the Skew-Detailed-Balance Markov chain of Eq. \ref{liftedproj} for the discrete magnetization $M$
by some Markov process for continuous $M$, one obtains the following run-and-tumble dynamics with velocities $(\pm N)$
\begin{eqnarray}
  \partial_t P_{t}^+( M )  &&  =  -\partial_M \left[ N P_{t}^+( M )\right] 
  -    {\tilde \Gamma}^+(M)  P_{t}^+( M )
 +   {\tilde \Gamma}^-(M)  P_{t}^-( M )
 \nonumber \\
   \partial_t P_{t}^-( M )  &&  =  \partial_M \left[ N P_{t}^+( M )\right] 
   -    {\tilde \Gamma}^+(M)  P_{t}^+( M )
 +   {\tilde \Gamma}^-(M)  P_{t}^-( M )
 \label{runMbig}
\end{eqnarray}
while the switching rates ${\tilde \Gamma}^{\pm}(M) $  involve the derivative $V'(m)$ discussed in Eq. \ref{vmderi} and the Heaviside $\theta$ function
\begin{eqnarray}
 {\tilde \Gamma}^+(M) && \equiv N \theta\left(V' \left(\frac{M}{N} \right) \right)   V' \left(\frac{M}{N} \right) 
 \nonumber \\
 {\tilde \Gamma}^-(M) && \equiv   N \theta\left(- V' \left(\frac{M}{N} \right) \right)  \left [- V' \left(\frac{M}{N} \right)  \right]   
\label{tildegamma}
\end{eqnarray}

For the probabilities $p^{\pm}_t(m)$ of the intensive magnetization $m=\frac{M}{N}$, 
the dynamics of Eqs \ref{runMbig}
 translate into the run-and-tumble process 
 involving the velocities $(\pm 1)$
\begin{eqnarray}
  \partial_t p_{t}^+( m )  &&  =  -\partial_m \left[  p_{t}^+( m )  \right] 
  -   \gamma^+ (m) p_{t}^+( m ) 
  +  \gamma^+ (m) p_{t}^-( m ) 
 \nonumber \\
   \partial_t p_{t}^-( m )  &&  =  \partial_m \left[  p_{t}^-( m ) \right] 
       +   \gamma^+ (m) p_{t}^+( m ) 
  -  \gamma^+ (m) p_{t}^-( m ) 
\label{runmsmall}
\end{eqnarray}
while the switching rates remain of order $N$
\begin{eqnarray}
\gamma^+ (m) && =  N  \theta\left(V' (m) \right)   V' (m) 
 \nonumber \\
\gamma^- (m) && =  N  \theta\left(-V' (m) \right)   \left[ -V'(m) \right]
\label{smallgamma}
\end{eqnarray}
Again to be more concrete, it is useful to write more explicitly the two cases :

(i) in the high-temperature phase $\beta J < \beta_c J=1$ or exactly at criticality,
the switching rates read (see Eq. \ref{vprimehigh})
\begin{eqnarray}
&&   \gamma^+(m) = 0  \ \ \ \ \ { \rm and } \ \ \ \  \   \gamma^-(m)  = -N V'(m)
\ \  \ \ \  {\rm for } \ \ m<0
  \nonumber \\  
&&  \gamma^+(m) = N V'(m)  \ \ \ \ \ \ { \rm and } \ \ \ \ \ \ \gamma^-(m) =0
\ \  \ \ \ {\rm for } \ \ m>0
\label{smallgammahigh}
\end{eqnarray}

(ii) in the low-temperature phase $\beta J > \beta_c J=1$,
the switching rates of Eq. \ref{gammathetaV} read (see Eq. \ref{vprime})
\begin{eqnarray}
&&  \gamma^+(m) = 0  \ \ \ \ \ { \rm and } \ \ \ \  \   \gamma^-(m)  = -N V'(m)
\ \  \ \ \ {\rm for } \ \ -1 \leq m< - m_{sp}(\beta) \ \ \  { \rm and } \ \ 0<m< m_{sp}(\beta)
  \nonumber \\  
&&   \gamma^+(m) = N V'(m)  \ \ \ \ \ \ { \rm and } \ \ \ \ \ \ \gamma^-(m) =0
\ \  \ \ \  {\rm for } \   - m_{sp}(\beta)<m<0 \ \ { \rm and } \ \ \ \  m_{sp}(\beta) <m \leq 1
\label{smallgammalow}
\end{eqnarray}

The steady state of Eq. \ref{runmsmall}
reads in terms of the equilibrium distribution of Eq. \ref{Peqm}
\begin{eqnarray}
  p_{eq}^{\pm}(m) = \frac{1}{2} \ \ \frac{ e^{-  N  V(m) } }  { \int_{-1}^{+1} dx e^{-  N  V(x) } }
\label{Peqmpm}
\end{eqnarray}
Here, the skew-detailed-balance property means that in the steady state, 
the positive current $j_{eq}^+(m)= p_{eq}^+( m ) >0 $ associated to the velocity $v=+1$ in the copy $(+)$
and the negative current $j_{eq}^-(m)= - p_{eq}^-( m ) <0 $ associated to the velocity $v=-1$ in the copy $(-)$
cancel each other to produce a vanishing total current
\begin{eqnarray}
 j_{eq}^{tot}(m) =j_{eq}^+(m) + j_{eq}^-(m) =0
\label{jtot}
\end{eqnarray}

The Green function governing the typical convergence properties (see the reminder in Appendix \ref{sec_green}) 
is computed in Appendix \ref{app_greenSDBrun},
while here we discuss the large deviations properties.


\subsection{ Explicit large deviations at Level 2.5 }

The large deviations at various levels have been analyzed recently
for the more general one-dimensional run-and-tumble processes with space-dependent velocities $v_{\pm}(m)$
and switching rates $\gamma_{\pm}(m)$ \cite{c_runandtumble}.
Let us summarize the main results for our present simpler model where the two velocities are unity $v_{\pm}(m)=1$,
while the two switching rates $\gamma_{\pm}(m)$ have disjoint supports,
starting with the relevant empirical observables.

(i) The empirical densities $ \rho^{\sigma}( m)  $ of the intensive magnetization $m$ 
in the two copies $\sigma=\pm$
\begin{eqnarray}
 \rho^{\sigma}( m) && \equiv \frac{1}{T} \int_0^T dt \  \delta_{\sigma(t),\sigma}  \ \delta (  m(t)-  m)  
\label{rhodiff}
\end{eqnarray}
satisfy the global normalization
\begin{eqnarray}
 \int_{-1}^{+1} dx \left[  \rho^+( m)  +  \rho^-( m) \right] =1
\label{rho1ptnormadiff}
\end{eqnarray}

(ii) The empirical magnetization currents in the two copies $\sigma=\pm 1$
\begin{eqnarray}
 j^{\sigma}( m) && \equiv \frac{1}{T} \int_0^T dt \ \frac{d m(t)}{dt}   \delta_{\sigma(t),\sigma}  \ \delta (  m(t)-  m)  
\label{jballdef}
\end{eqnarray}
are directly given by 
the empirical densities $ \rho^{\sigma}( m)  $ as a consequence of the ballistic motion at velocity unity
in each copy
\begin{eqnarray}
 j^+(m) &&=  \rho^{+}( m) 
 \nonumber \\
  j^-(m) &&=  - \rho^{-}( m) 
\label{jballistic}
\end{eqnarray}

(iii)
The jumps between the two internal states $ \sigma=\pm$ at magnetization $m$ are described by the empirical switching flows
\begin{eqnarray}
Q^{+}( m)  \equiv  \frac{1}{T} \sum_{t  \in [0,T]: \sigma(t^+) \ne \sigma(t^-) } \delta_{\sigma(t^-),+} \  \delta(  m(t)-  m) 
\nonumber \\
Q^{-}( m)  \equiv  \frac{1}{T} \sum_{t  \in [0,T]: \sigma(t^+) \ne \sigma(t^-) } \delta_{\sigma(t^-),-} \  \delta(  m(t)-  m)
\label{jumpflowsm}
\end{eqnarray}
The stationarity constraints read
\begin{eqnarray}
0 && = -   \frac{d j^+( m)  }{dm} - Q^{+}( m)  + Q^{-}( m)  
\nonumber \\
0 && =  -  \frac{d j^-( m)  }{dm}+  Q^{+}( m)  - Q^{-}( m)  
\label{jumpstatio}
\end{eqnarray}
The sum yields that the total magnetization current $ \left[  j^+( m)+ j^-( m)\right] $ cannot depend on $m$ and thus has to vanish, as a consequence of the boundary conditions 
at $m= \pm 1$ without flows.
So the two empirical densities coincide, and will be rewritten in terms of the total density
 $\rho(m)$ from now on
\begin{eqnarray}
 \rho^+(m) =  \rho^{-}( m)  = \frac{\rho(m)}{2}
\label{jxtotempi}
\end{eqnarray}
The remaining stationarity constraint reads
\begin{eqnarray}
\frac{ \rho'( m)}{2}  =  - Q^{+}( m)  + Q^{-}( m)  
\label{statio}
\end{eqnarray}
For each magnetization $m$, one of the two switching flows $Q^{\pm}(m) $ does not exist,
since one of the two switching rates $\gamma^{\pm}(M)$ vanishes as a consequence of the choice of Eq. \ref{smallgamma}.
So the stationarity condition of Eq. \ref{statio}
can be rewritten more explicitly as
\begin{eqnarray}
   Q^{+}( m)=   - \frac{ \rho'( m)}{2}        \ \ \ {\rm if } \ \ V'(m) >0
    \nonumber \\
     Q^{-}( m)=    \frac{ \rho'( m)}{2}      \ \ \ {\rm if } \ \ V'(m) <0 
 \label{elimQjpm}
\end{eqnarray}
Putting everything together, the large deviations at Level 2.5
read for the joint probability to see the total empirical density $\rho(m)$
and the switching flows $Q^{\pm}(m)$
\begin{eqnarray}
 P_T^{Level \ 2.5}[ \rho(.) , Q_{\pm}(.) ]  && \opsimeq_{T \to +\infty} 
  e^{- \displaystyle T  I_{2.5}[\rho(.),  Q_{\pm}(.) ]   }
  \delta \left( \int_{-1}^{+1} dm \rho(m)   - 1  \right)
\nonumber \\
&&  \left[ \prod_{ m : V'(m) >0 }  \delta \left( Q^{+}( m)   + \frac{\rho'(m) }{2}    \right)  \right] 
 \left[ \prod_{ m : V'(m) <0 }  \delta \left( Q^{-}( m)   - \frac{\rho'(m) }{2}    \right)  \right] 
\label{ld2.5rhoqm}
\end{eqnarray}
with the rate function
\begin{eqnarray}
 I_{2.5}[ \rho(.),  Q_{\pm}(.) ]  && = 
  \int_{-1}^{+1}d  m \theta \left( V'(m) 0\right) 
\left[  Q^{+}( m)    \ln \left( \frac{  Q^{+}( m)   }{     N V' (m)  \frac{ \rho( m)}{2} }  \right) 
 -   Q^{+}( m)   +     N V' (m)  \frac{ \rho( m)}{2}    \right]  
\nonumber \\
&& +
   \int_{-1}^{+1}d  m \theta \left( -V'(m)\right) 
\left[  Q^{-}( m)    \ln \left( \frac{  Q^{-}( m)   }{    \left[ -N V'(m) \right] \frac{ \rho( m)}{2} }  \right) 
 -   Q^{-}( m)   +   \left[ -N V'(m) \right] \frac{ \rho( m)}{2}    \right]  
\label{rate2.5rhoqm}
\end{eqnarray}


\subsection{ Explicit large deviations at Level 2 for the empirical density $ \rho(m)$ alone}

One can use the constraints on the second line of Eq. \ref{ld2.5rhoqm}
to eliminate the switching flows $Q^{\pm}(m)$
and one obtains the large deviations at Level 2 for the empirical density $ \rho(m)$ alone
\begin{eqnarray}
 P_T^{Level \ 2}[ \rho(.)  ]   \opsimeq_{T \to +\infty}   \delta \left( \int_{-1}^{+1} dm \rho(m)   - 1  \right)
 e^{- \displaystyle T  I_{2}[\rho(.)   }
\label{ld2rhoqm}
\end{eqnarray}
where the rate function at Level 2 reads
\begin{eqnarray}
 I_{2}[ \rho(.) ]  && = 
\frac{1}{2}  \int_{-1}^{+1}d  m \ \theta\left( V'(m) \right) \theta\left( -\rho'(m) \right)
\left[   -  \rho'( m)     \ln \left( \frac{   -  \rho'( m)   }{     N V'(m)  \rho( m)}  \right) 
  +  \rho'( m)    +    N V'(m) \rho( m)    \right]  
\nonumber \\
&& +
\frac{1}{2}  \int_{-1}^{+1}d  m \ \theta( - V'(m) )\theta\left( \rho'(m) \right)
\left[  \rho'( m)    \ln \left( \frac{   \rho'( m)  }{    -  N V'(m)  \rho( m) }  \right) 
 -    \rho'( m)    - N V'(m) \rho( m)  \right]  
 \nonumber \\
&& = 
\frac{N}{2}  \int_{-1}^{+1}d  m \ \theta\left( V'(m) \right) \theta\left( -\rho'(m) \right)
\rho(m) 
\left[   -   \frac{ \rho'( m) }{ N \rho(m) }     \ln \left( \frac{   -  \rho'( m)   }{     N V'(m)  \rho( m)}  \right) 
  +   \frac{ \rho'( m) }{ N \rho(m) }    +     V'(m)     \right]  
\nonumber \\
&& +
\frac{N}{2}  \int_{-1}^{+1}d  m \ \theta( - V'(m) ) \theta\left( \rho'(m) \right)
\rho(m) 
\left[  \frac{ \rho'( m) }{ N \rho(m) }   \ln \left( \frac{   \rho'( m)  }{    -  N V'(m)  \rho( m) }  \right) 
 -    \frac{ \rho'( m) }{ N \rho(m) }    -  V'(m)   \right]   
\label{rate2runtheta}
\end{eqnarray}
Again it is again convenient to parametrize the normalized empirical density $\rho(.)$ via the empirical potential $U(.)$
of Eq. \ref{rhomparaU} to rewrite the rate function of Eq. \ref{rate2runtheta} as
\begin{eqnarray}
&& I_{2}^{SDB} \left[ \rho(.)=\frac{ e^{-  N  U(.) } }  { \int_{-1}^{+1} dx e^{-  N  U(x) } } \right] 
 \nonumber \\
&&    = 
\frac{ N  \int_{-1}^{+1}d  m \ e^{-  N  U(m) } \ 
\theta\left( V'(m) \right) \theta\left( U'(m) \right)  
\left[   U'(m)   \ln \left( \frac{   U'( m)   }{      V'(m)  }  \right)   - U'(m)   +     V'(m)     \right] } 
 { 2 \int_{-1}^{+1} dx e^{-  N  U(x) } } 
\nonumber \\
&& +
 \frac{ N  \int_{-1}^{+1}d  m \ e^{-  N  U(m) } \ \theta( - V'(m) ) \theta\left( - U'(m) \right) 
 \left[  -U'(m)  \ln \left( \frac{   U'( m)  }{    V'(m)  }  \right)  +U'(m)   -  V'(m)   \right]    }
   { 2 \int_{-1}^{+1} dx e^{-  N  U(x) } }
\label{rate2runthetapara}
\end{eqnarray}
that should be compared with its detailed-balance counterpart of Eq. \ref{rate2diffUpara}.
In particular, the cost of a small perturbation $\epsilon(m)=U(m)-V(m)$ of the empirical potential $U(m)$ around the equilibrium potential $V(m)$ 
\begin{eqnarray}
 I_{2}^{SDB} \left[ \rho(.)=\frac{ e^{-  N  \left[ V(.) + \epsilon(.) \right] } }  { \int_{-1}^{+1} dx e^{-  N  \left[ V(x) + \epsilon(x) \right] } } \right] 
 \opsimeq_{\epsilon(.) \to 0}
\frac{ N  \int_{-1}^{+1}d  m \ e^{-  N  V(m) } \ 
 \frac{  [\epsilon'(m) ]^2   }{    2 \vert  V'(m) \vert }   } 
 { 2 \int_{-1}^{+1} dx e^{-  N  V(x) } } 
\label{rate2runthetaparaperturb}
\end{eqnarray}
involves again $[\epsilon'(m) ]^2 $ at lowest order as in Eq. \ref{rate2diffUparaperturb}, 
but the additional factor $\frac{ 1   }{    2 \vert  V'(m) \vert }     $ in the integral of the
numerator yields that the difference between Eq. \ref{rate2runthetaparaperturb}
and Eq. \ref{rate2diffUparaperturb}
does not have a definite sign.

As a final remark, let us mention that in the high-temperature phase and at criticality
where the switching rates are given by Eq. \ref{smallgammahigh},
further results can be found in Appendix D of \cite{c_runandtumble},
in particular for the decomposition into excursions around the origin.


\section{ Conclusion }

\label{sec_conclusion}

In this paper, we have revisited the Skew-Detailed-Balance Lifted-Markov-chain algorithm for the Curie-Weiss 
mean-field ferromagnetic model
  \cite{TCV,weigel,SDBcurierun,vuceljaa}.
  The goal was to study the large deviations at various levels for empirical time-averaged observables,
in order to compare with their detailed-balance counterparts.
  We have first considered the single-spin-flip dynamics for the discrete extensive magnetization $M$,
  where the detailed-balance dynamics corresponds to some non-directed one-dimensional Markov jump process,
  while the skew-detailed-balance dynamics corresponds to some directed Markov jump process on a ladder.
  We have then focused on the corresponding effective dynamics for the continuous intensive magnetization $m=\frac{M}{N}$ for large system-size $N$, 
where the detailed-balance dynamics corresponds to some one-dimensional Fokker-Planck dynamics,
while the skew-detailed-balance dynamics corresponds to some one-dimensional run-and-tumble process.

As in previous studies concerning various other non-reversible algorithms
 \cite{rey_langevin,rey_graph,rey_general,jack_accel,chetrite_sampling},
our main conclusion is that all the progresses achieved recently in the field of non-equilibrium steady states (see the
reminder and references in the Introduction)
are very useful to better understand the large deviations properties
 of empirical time-averaged observables in Monte-Carlo algorithms
breaking detailed-balance.


\appendix


\section{ Reminder on the Green function $G$ governing the typical convergence properties   }

\label{sec_green}

In this Appendix, we recall how the typical convergence properties of the Markov jump process of Eq. \ref{master}
are governed by the Green function $G$,
as already emphasized for the Fokker-Planck dynamics in the context of Anderson localization \cite{ramola}, 
and for Markov chains either in discrete time or in continuous time 
in the context of Ruelle thermodynamic formalism for Markov trajectories
 \cite{c_ruelle}.


\subsection{ Convergence of the finite-time propagator $\langle x \vert e^{wt} \vert x_0 \rangle $ towards the steady state $P_{*}(x) $ }

For the continuous-time Markov Chain of generator $w$ of Eq. \ref{master},
the difference between the finite-time propagator $\langle x \vert e^{wt} \vert x_0 \rangle $ and the steady state $P_{*}(x) $
\begin{eqnarray}
 \Delta_t(x,x_0)    \equiv \langle x \vert e^{wt} \vert x_0 \rangle - P_{*}(x)
 = \langle x \vert e^{wt} \vert x_0 \rangle - \langle x \vert r_0 \rangle \langle l_0 \vert x_0 \rangle
 =  \langle x \vert \left[ e^{wt} - \vert r_0 \rangle \langle l_0 \vert\right] \vert x_0 \rangle 
\label{defdelta}
\end{eqnarray}
decays towards zero for large time $ t \to +\infty$
\begin{eqnarray}
 \Delta_t(x,x_0)    \operarrow_{t \to +\infty} 0
\label{deltainfty}
\end{eqnarray}
starting from the initial condition at $t=0$
\begin{eqnarray}
 \Delta_{t=0}(x,x_0)   = \delta_{ x,x_0} - P_{*}(x)
\label{deltaini}
\end{eqnarray}
while remaining orthogonal to the left eigenvector $\langle l_0 \vert  $ and to the right eigenvector $ \vert r_0 \rangle$
 \begin{eqnarray}
  \langle l_0 \vert  \Delta_t && = \sum_x l_0(x)  \Delta_t(x,x_0) =  \sum_x  \Delta_t(x,x_0)  = 0
 \nonumber \\
    \Delta_t \vert r_0 \rangle && =  \sum_{x_0}   \Delta_t(x,x_0) r_0(x_0)=  \sum_{x_0}   \Delta_t(x,x_0) P_*(x_0)
    = 0 
\label{deltaorthog}
\end{eqnarray}
Its dynamics can be written as the forward master equation with respect to the final position $x$
or as the backwards master equation with respect to the initial position $x_0$
\begin{eqnarray}
\partial_t \Delta_t(x,x_0) &&  \!\!\!\!\!  = \langle x \vert w  e^{wt} \vert x_0 \rangle 
= \sum_y w(x,y) \langle y \vert   e^{wt} \vert x_0 \rangle
= \sum_y w(x,y) \left[ \Delta_t(y,x_0) + P_*(y) \right] = \sum_y w(x,y)  \Delta_t(y,x_0) 
\label{dyndelta}
 \\
\partial_t \Delta_t(x,x_0) && \!\!\!\!\! =  \langle x \vert   e^{wt} w \vert x_0 \rangle
= \sum_{y_0}  \langle x \vert   e^{wt} \vert y_0 \rangle w(y_0,x_0)
= \sum_{y_0} \left[ \Delta_t(x,y_0) + P_*(x) \right]  w(y_0,x_0)
= \sum_{y_0}  \Delta_t(x,y_0)  w(y_0,x_0)
 \nonumber
\end{eqnarray}


\subsection{ Inverse $G$ of the generator $w$ in the subspace orthogonal to the zero-eigenvalue-subspace of $w$ }

\label{subsec_Ginverse}

As will be explained in detail in further subsections, 
in order to characterize the convergence properties of empirical
time-averaged observables, it is not necessary to analyze the full time-dependence of 
the difference $ \Delta_t(x,x_0) $ of Eq. \ref{defdelta}, and
it is actually sufficient to focus on the Green function $G(x,x_0) $
defined as the integral over time $ t \in [0,+\infty[$ of the difference $ \Delta_t(x,x_0) $ 
\begin{eqnarray}
G(x,x_0) \equiv \int_0^{+\infty} dt \Delta_t(x,x_0)  
\label{defgreen}
\end{eqnarray}
The integral over time of the dynamical Eqs \ref{dyndelta} for the difference $ \Delta_t(x,x_0) $ 
and the boundary conditions at $t=+\infty$ (Eq. \ref{deltainfty}) and at $t=0$ (Eq. \ref{deltaini}) yield the forward and the backward equations
\begin{eqnarray}
\sum_y w(x,y)  G(y,x_0) && =\int_0^{+\infty} dt \left[ \sum_y w(x,y)  \Delta_t(y,x_0) \right] =
\int_0^{+\infty} dt \partial_t \Delta_t(x,x_0) = \Delta_{t=+\infty}(x,x_0) - \Delta_{t=0}(x,x_0) 
\nonumber \\
&& =P_{*}(x) - \delta_{ x,x_0}  
 \nonumber \\
\sum_{y_0}  G(x,y_0)  w(y_0,x_0) && =\int_0^{+\infty} dt \left[ \sum_{y_0}  \Delta_t(x,y_0)  w(y_0,x_0) \right] =
\int_0^{+\infty} dt\partial_t \Delta_t(x,x_0) =  \Delta_{t=+\infty}(x,x_0) - \Delta_{t=0}(x,x_0) 
\nonumber \\
&& =P_{*}(x) - \delta_{ x,x_0}  
 \label{eqG}
\end{eqnarray}
At the matrix level, these two equations can be rewritten as
\begin{eqnarray}
 (-w)   G &&  =  \mathbb{1} - \vert r_0 \rangle \langle l_0 \vert  
 \nonumber \\
G (-w)  && =  \mathbb{1} - \vert r_0 \rangle \langle l_0 \vert  
 \label{eqGop}
\end{eqnarray}
while Eqs \ref{deltaorthog} yield the orthogonality of the matrix $G$ 
with the left eigenvector $\langle l_0 \vert  $ and with the right eigenvector $ \vert r_0 \rangle$
 \begin{eqnarray}
0=  \langle l_0 \vert  G && = \sum_x l_0(x) G(x,x_0) =  \sum_x  G(x,x_0)  
 \nonumber \\
0=    G \vert r_0 \rangle && =  \sum_{x_0}   G(x,x_0) r_0(x_0)=  \sum_{x_0}   G(x,x_0) P_*(x_0) 
\label{Gorthog}
\end{eqnarray}
In conclusion, the matrix $G$ is the inverse of the matrix $(-w)$  within the subspace $\left( \mathbb{1} - \vert r_0 \rangle \langle l_0 \vert   \right) $ orthogonal to 
the subspace $\left(  \vert r_0 \rangle \langle l_0 \vert   \right) $ associated to the zero eigenvalue of $w$
\begin{eqnarray}
   G  = \left( \mathbb{1} - \vert r_0 \rangle \langle l_0 \vert  \right) \frac{1}{(-w) }  \left( \mathbb{1} - \vert r_0 \rangle \langle l_0 \vert  \right)
 \label{Gasinverse}
\end{eqnarray}
This Green function is of course familiar from the perturbation theory of isolated eigenvalues 
(see the subsection \ref{subsec_per} of the main text).


\subsection{ Explicit computation of the Green function $G$ in specific models }

\label{subsec_practice}

In specific models that are simple enough (see the four examples in the next four appendices), the Green function $G(x,x_0)$ 
can be computed a priori via two methods :

(i) either one considers that the initial position $x_0$ is a parameter, and the function $G(.,x_0) $
can be found as the solution of the forward equation of Eq. \ref{eqG} with respect to the final position,
that reads using Eq. \ref{wdiag}
\begin{eqnarray}
P_{*}(x) - \delta_{ x,x_0}  = \sum_y w(x,y)  G(y,x_0) = \sum_{ y \ne x} \left[ w(x,y)  G(y,x_0) -  w(y,x)  G(x,x_0) \right]
 \label{Ginvright}
\end{eqnarray}
and then one needs to impose the first orthogonality condition of Eq. \ref{Gorthog}
 \begin{eqnarray}
0=    \sum_x  G(x,x_0)  
\label{Gorthogright}
\end{eqnarray}

(ii) or one considers instead that the final position $x$ is a parameter, and the function $G(x,.) $ 
can be found as the solution of the backward equation of Eq. \ref{eqG} 
with respect to the initial position,
that reads using Eq. \ref{wdiag}
\begin{eqnarray}
P_{*}(x) - \delta_{ x,x_0}  = \sum_{y_0}  G(x,y_0)  w(y_0,x_0) 
=  \sum_{y_0 \ne x_0}  \left[ G(x,y_0) - G(x,x_0) \right] w(y_0,x_0) 
 \label{Ginvleft}
\end{eqnarray}
and then one needs to impose the second orthogonality condition of \ref{Gorthog}
 \begin{eqnarray}
0=      \sum_{x_0}   G(x,x_0) P_*(x_0) = G(x,x) +   \sum_{x_0 \ne x }  \left[ G(x,x_0) - G(x,x) \right]P_*(x_0) 
\label{Gorthogleft}
\end{eqnarray}

In practice, the method (ii) is technically simpler 
because the backward Eq. \ref{Ginvleft} only involves differences 
$\left[ G(x,y_0) - G(x,x_0) \right] $ of the Green function
between points $y_0 \ne x_0$ that are connected via transitions rates $w(y_0,x_0) >0 $ of the Markov matrix $w$,
and because the left hand side of Eq. \ref{Ginvleft} only involves the delta function $ \delta_{ x,x_0} $ with respect to 
the variable $x_0$, while the other term $P_{*}(x)  $ does not depend on $x_0$.
Then Eq. \ref{Gorthogleft} allows to determine the Green function $G(x,x)$ at coinciding points
in terms of the differences $ \left[ G(x,x_0) - G(x,x) \right] $ computed from the solution of Eq. \ref{Ginvleft}.
The fact that the method (ii) is simpler can be also understood via the link
with the Mean-First-Passage-Time $\tau(x,x_0)$ as we now explain.


\subsection{ Link with the Mean-First-Passage-Time $\tau(x,x_0)$ at position $x$ when starting at position $x_0$ }

\label{subsec_firstpassage}

The probability distribution $F(t_1;x,x_0)$ of the first-passage-time $t_1$ at position $x$ when starting at $x_0$
with the normalization 
 \begin{eqnarray}
\int_0^{+\infty} dt_1 F(t_1;x,x_0) =1
\label{firstpassagedistri}
\end{eqnarray}
allows to compute the Mean-First-Passage-Time $\tau(x,x_0)$ at position
$x$ when starting at $x_0$
 \begin{eqnarray}
\tau(x,x_0) \equiv \int_0^{+\infty} dt_1 t_1 F(t_1;x,x_0)
\label{mfpt}
\end{eqnarray}
that of course vanishes at coinciding points $x_0=x$
 \begin{eqnarray}
\tau(x,x) =0
\label{mfptcoinciding}
\end{eqnarray}

The finite-time propagator $\langle x \vert e^{wt} \vert x_0 \rangle $ can be rewritten in terms of the
probability distribution $F(t_1;x,x_0)$ of the first-passage-time $t_1$ via
the convolution with the propagator $\langle x \vert e^{w(t-t_1)} \vert x \rangle $ at coinciding points for the remaining available time $(t-t_1)$
\begin{eqnarray}
\langle x \vert e^{wt} \vert x_0 \rangle = \int_0^t dt_1 \langle x \vert e^{w (t-t_1) } \vert x \rangle F(t_1;x,x_0)
\label{propagatorandfirst}
\end{eqnarray}
The rewriting of this identity in terms of the difference $ \Delta_t(x,x_0) $ of Eq. \ref{defdelta}
reads using the normalization of Eq. \ref{firstpassagedistri}
\begin{eqnarray}
 \Delta_t(x,x_0)  &&  = - P_{*}(x)  + \int_0^t dt_1 \left[ P_{*}(x) + \Delta_{t-t_1}(x,x)  \right]   F(t_1;x,x_0)
 \nonumber \\
&& =  - P_{*}(x) \int_t^{+\infty} dt_1 F(t_1;x,x_0) + \int_0^t dt_1  \Delta_{t-t_1}(x,x)    F(t_1;x,x_0)
\label{deltaetfirst}
\end{eqnarray}
The integration over time $t \in [0,+\infty[$ yields for the Green function $G(x,x_0) $ of Eq. \ref{defgreen}
using the normalization of Eq \ref{firstpassagedistri}
and the definition of the Mean-First-Passage-Time $\tau(x,x_0)$ of Eq. \ref{mfpt}
\begin{eqnarray}
G(x,x_0) && = \int_0^{+\infty} dt \Delta_t(x,x_0)  
=  - P_{*}(x) \int_0^{+\infty} dt \int_t^{+\infty} dt_1 F(t_1;x,x_0) + \int_0^{+\infty} dt \int_0^t dt_1  \Delta_{t-t_1}(x,x)    F(t_1;x,x_0)
\nonumber \\
&& =  - P_{*}(x)  \int_0^{+\infty} dt_1 t_1 F(t_1;x,x_0) + \int_0^{+\infty} dt_2 \Delta_{t_2}(x,x) \int_0^{+\infty} dt_1      F(t_1;x,x_0)
\nonumber \\
&& =  - P_{*}(x) \tau(x,x_0)  + G(x,x) 
\label{greenandfirst}
\end{eqnarray}
So the dependence upon the initial position $x_0$ of the Green function $G(x,x_0) $ is only via the MFPT $\tau(x,x_0)$ of Eq. \ref{mfpt}. Let us now discuss various consequences :

(a) Plugging Eq. \ref{greenandfirst} into the backward Eq. \ref{Ginvleft} for Green function $G(x,x_0) $
yields the following backward equation for the MFPT $\tau(x,x_0)$
\begin{eqnarray}
-1 + \frac{ \delta_{ x,x_0}  }{ P_*(x)} =  \sum_{y_0 \ne x_0}  \left[   \tau(x,y_0)  - \tau(x,x_0) \right] w(y_0,x_0) = 
 \sum_{y_0 }  \tau(x,y_0)   w(y_0,x_0) 
 \label{Ginvlefttau}
\end{eqnarray}
 This backward equation is of course very well-known for $x_0 \ne x$,
 and is a standard method to compute the MFPT $\tau(x,x_0)$ with the boundary condition
 of Eq. \ref{mfptcoinciding} for Markov processes (see for instance the textbooks \cite{gardiner,vankampen,risken,redner}).
 Eq. \ref{Ginvlefttau} for $x=x_0$ is usually not written because it is not needed to compute the MFPT,
 but the delta function term is nevertheless necessary for consistency when one projects the equation onto $\vert r_0 \rangle$,
 i.e. when one multiplies the equation by $  P_*(x_0)$ and then sum over $x_0$.

(b) Plugging Eq. \ref{greenandfirst}
 into the orthogonality condition of Eq. \ref{Gorthogleft} yields that the
 Green function $G(x,x) $ at coinciding points $x_0=x$
 \begin{eqnarray}
 G(x,x) = -   \sum_{x_0 \ne x }  \left[ G(x,x_0) - G(x,x) \right]P_*(x_0) =     P_{*}(x)   \sum_{x_0 \ne x } \tau(x,x_0) P_*(x_0)
 = P_{*}(x)   \sum_{x_0  } \tau(x,x_0) P_*(x_0)
\label{Gorthoglefttau}
\end{eqnarray}
is directly related to the average of the MFPT $\tau(x,x_0)$ over the initial condition $x_0$ drawn with the steady-state distribution
$ P_*(x_0)$

(c)
Plugging Eq. \ref{greenandfirst}
 into the orthogonality condition of Eq. \ref{Gorthogright} valid for any $x_0$
 \begin{eqnarray}
0=    \sum_x  G(x,x_0)  = 
 -  \sum_x  P_{*}(x) \tau(x,x_0)   +    \sum_x \sum_{y_0 } P_{*}(x) \tau(x,y_0)  P_*(y_0) 
\label{Gorthogrighttau}
\end{eqnarray}
shows that the average of the MFPT $ \tau(x,x_0)$ over the final point $x$ drawn with the steady state distribution $ P_*(x)$
is actually independent of the initial point $x_0$
 \begin{eqnarray}
\tau_{eq}(x_0) \equiv  \sum_x P_{*}(x) \tau(x,x_0) = \tau_{eq} \ \ \ \ { \rm independent \ \ of } \ \ x_0
\label{taueqx0indeo}
\end{eqnarray}
and that its value coincides with the trace of the Green matrix,
i.e. to the sum of the diagonal elements $G(x,x)$ of Eq. \ref{Gorthoglefttau}
 \begin{eqnarray}
\tau_{eq}  =  \sum_x \sum_{x_0 } P_{*}(x) \tau(x,x_0)  P_*(x_0) = \sum_x G(x,x)
\label{taueqtraceg}
\end{eqnarray}
This constant is known in the mathematical literature as Kemeny’s constant 
(see \cite{Kemeny_doyle,Kemeny_hunter,Kemeny_bini,Kemeny1d} and references therein).


\subsection{ Interpretation of the Green function $G$ in terms of the relaxation spectrum when $w$ is diagonalizable }

Whenever the Markov matrix $w$ can be diagonalized 
\begin{eqnarray}
 w  = -  \sum_{n>0} \zeta_n \vert r_n \rangle \langle l_n  \vert 
\label{jumpspectral}
\end{eqnarray}
in terms of the other eigenvalues $(-\zeta_n)<0$ labelled by $n>0$,
with their right eigenvectors $\vert r_n \rangle $ and their left eigenvectors $ \langle l_n  \vert  $
\begin{eqnarray}
 w \vert r_n \rangle && = -   \zeta_n \vert r_n \rangle 
 \nonumber \\
 \langle l_n  \vert  w  && = -   \zeta_n  \langle l_n  \vert 
\label{defeigenzetan}
\end{eqnarray}
satisfying the othonormalization and closure relations
\begin{eqnarray}
 \langle l_n  \vert r_m \rangle && = \delta_{nm}
 \nonumber \\
 \mathbb{1} && = \vert r_0 \rangle \langle l_0 \vert + \sum_{n>0}  \vert r_n \rangle \langle l_n  \vert
\label{fermeturej}
\end{eqnarray}
the finite-time generator can be decomposed into these relaxation modes $n>0$ towards
the steady state $n=0$
\begin{eqnarray}
 e^{wt }  = \vert r_0 \rangle \langle l_0  \vert  +  \sum_{n>0} e^{- t \zeta_n } \vert r_n \rangle \langle l_n  \vert 
\label{expwspectral}
\end{eqnarray}
As a consequence,
the difference $\Delta_t$ of Eq. \ref{defdelta} reads
\begin{eqnarray}
\Delta_t(x,x_0) =  \sum_{n>0} e^{  -  t \zeta_n }
\langle x \vert r_n \rangle \langle l_n  \vert x_0 \rangle
=  \sum_{n>0} e^{  - t \zeta_n } r_n(x) l_n(x_0)
\label{jumppropa}
\end{eqnarray}
so that the Green function of Eq. \ref{defgreen} 
\begin{eqnarray}
G(x,x_0) = \int_0^{+\infty} dt \Delta_t(x,x_0)  =  \sum_{n>0} \frac{  r_n(x) l_n(x_0) }{\zeta_n }
\label{greenrelax}
\end{eqnarray}
represents some global information on the relaxation between the two points $x_0$ and $x$.
The MFPT $\tau(x,x_0)$ then reads from Eq. \ref{greenandfirst}
\begin{eqnarray}
 \tau(x,x_0)  = \frac{G(x,x) -G(x,x_0)}{P_*(x)}
= \sum_{n>0} \left[\frac{r_n(x)}{r_0(x) } \right] \frac{ 1 }{\zeta_n } \left[ l_n(x) - l_n(x_0)\right]
\label{mfptrelax}
\end{eqnarray}
This spectral decomposition allows to recover the backward equation of Eq. \ref{Ginvlefttau} 
from the properties of eigenvectors, as it should for consistency.
Then the equilibrium time of Eq. \ref{taueqx0indeo} can be computed using the properties of the eigenvectors
 \begin{eqnarray}
\tau_{eq}(x_0) && \equiv  \sum_x P_{*}(x) \tau(x,x_0) 
=\sum_{n>0} \frac{  1 }{\zeta_n } \left[ \sum_x r_n(x)  l_n(x) -  l_n(x_0) \sum_x r_n(x) \right]
 = \sum_{n>0} \frac{  1 }{\zeta_n } \left[ \langle l_n \vert r_n \rangle -  l_n(x_0)\langle l_0 \vert r_n \rangle  \right]
 \nonumber \\
 && = \sum_{n>0} \frac{  1 }{\zeta_n } \equiv \tau_{eq}
\label{taueqx0indeosp}
\end{eqnarray}
i.e. one recovers that it does not depend upon the initial position $x_0$ (Eq. \ref{taueqx0indeo})
and that its value $\tau_{eq} $ reducing to the sum of the inverses of the relaxation eigenvalues $\zeta_n$
indeed coincides with the trace of the Green matrix,
i.e. to the sum of the diagonal elements $G(x,x)$ of Eq. \ref{greenrelax}
 \begin{eqnarray}
 \sum_x G(x,x) =  \sum_{n>0} \frac{ 1 }{\zeta_n }  \sum_x  r_n(x) l_n(x)
 =  \sum_{n>0} \frac{ \langle l_n \vert r_n \rangle  }{\zeta_n }
 = \sum_{n>0} \frac{  1 }{\zeta_n } = \tau_{eq}
\label{taueqtracegsp}
\end{eqnarray}

In summary, these spectral decompositions are 
very nice to better understand the physical meaning of the Green function $G(x,x_0) $,
 and of the MFPT $\tau(x,x_0)$,
but it is important to stress the two following points :

(i) Even when the Markov matrix $w$ is diagonalisable, one does not need to be able to compute the whole relaxation spectrum to obtain the Green function $G$ via its spectral decomposition 
of Eq. \ref{greenrelax}, since the Green function can be obtained directly via the much simpler method
discussed in the subsection \ref{subsec_practice}.

(ii) When the Markov matrix $w$ is not diagonalisable,
the Green function defined by Eq. \ref{defgreen} 
characterizes nevertheless the convergence properties
towards the steady state as explained in subsection \ref{subsec_Ginverse},
and can be computed again computed via the methods of subsection \ref{subsec_practice}.

In the following subsections, we describe how the Green function $G$ governs
the typical convergence properties of empirical time-averaged observables.
 

\subsection{ Convergence of the averaged value $< \rho_T(x) >_{Traj} $ of the empirical density towards 
the steady state $P_*(x) $ }

The averaged value of the empirical density $\rho_T( x )  $ of Eq. \ref{rhoc}
over the dynamical trajectories 
can be rewritten in terms of the finite-time propagator $\langle x \vert e^{wt} \vert x_0 \rangle $ 
\begin{eqnarray}
<  \rho_T(x) >_{Traj}  =   \frac{1}{T} \int_0^{T} dt < \delta_{x(t) , x } >_{Traj} 
  = \frac{1}{T} \int_0^{T} dt  \langle x \vert e^{wt} \vert x_0 \rangle
\label{avadditive}
\end{eqnarray}
and thus in terms of the difference $\Delta_t(x,x_0)$ of Eq. \ref{defdelta}
to show explicitly the convergence towards the steady state $ P_*(x)$
\begin{eqnarray}
< \rho_T(x) >_{Traj} 
  =   \frac{1}{T} \int_0^{T} dt \left[ P_*(x) + \Delta_t(x,x_0)  \right]
  =  P_*(x)+  \frac{1}{T}  \int_0^{T} dt  \Delta_t(x,x_0) 
\label{avadditivespectral}
\end{eqnarray}
So the leading correction of order $1/T$ involves the Green function $ G(x,x_0)$ of Eq. \ref{defgreen}
\begin{eqnarray}
< \rho_T(x) >_{Traj} -     P_*(x)\opsimeq_{T \to +\infty}   \frac{1}{T}  \int_0^{+\infty} dt  \Delta_t(x,x_0) 
= \frac{G(x,x_0)}{T}
\label{cvrhoav}
\end{eqnarray}


\subsection{ Convergence of empirical time-averaged observables ${\cal O}_T $
 towards their steady values $O_*$ } 
 
\label{subsec_empiObser}

Any empirical time-averaged observable of the form of Eq. \ref{defadditive}
can be rewritten in terms of the empirical density of Eq. \ref{rhoc}
\begin{eqnarray}
{\cal O}_T  = \sum_x O(x) \rho_T(x) 
\label{additiverho}
\end{eqnarray} 
As a consequence, the averaged value of the empirical observable ${\cal O}_T $
over the dynamical trajectories $[x(0 \leq t \leq T) ] $ 
involves the averaged value $< \rho_T(x) >_{Traj} $ of the empirical density discussed in Eq. \ref{cvrhoav}
\begin{eqnarray}
< {\cal O}_T >_{Traj} && = \sum_x O(x) < \rho_T(x) >_{Traj}
\opsimeq_{T \to +\infty}  \sum_x O(x) \left[  P_*(x) + \frac{G(x,x_0)}{T} \right]
\label{avArho}
\end{eqnarray} 
So the convergence towards the steady value of Eq. \ref{aeq}
is governed by the Green function $G(x,x_0) $ of Eq. \ref{defgreen}
\begin{eqnarray}
< {\cal O}_T >_{Traj}  - O_* && 
\opsimeq_{T \to +\infty} \frac{1}{T} \sum_x O(x) G(x,x_0)
\label{avArhocv}
\end{eqnarray}

The scaled variance of the empirical time-averaged observable of Eq. \ref{additiverho} 
can be rewritten in terms of propagators
\begin{eqnarray}
V_{ [{\cal O}_T] } && \equiv T \left( < {\cal O}^2_T >_{Traj} - \left(< {\cal O}_T >_{Traj} \right)^2 \right)
\nonumber \\
&& 
 =  \sum_{x} O(x)  \frac{1}{T} \sum_y O(y)  \int_0^T dt' \int_0^T dt
 \left( < \delta_{x(t') , x } \delta_{x(t) , y }  >_{Traj} - < \delta_{x(t') , x }>_{Traj} <  \delta_{x(t) , y }  >_{Traj} \right)
 \nonumber \\
&& 
 =  \sum_{x} O(x)   \sum_y O(y)
 \frac{2}{T}  \int_0^T dt  \int_0^{T-t} d\tau   
 \left( < \delta_{x(t+\tau) , x } \delta_{x(t) , y }  >_{Traj} - <  \delta_{x(t+\tau) , x }>_{Traj} <  \delta_{x(t) , y }  >_{Traj} \right) 
  \nonumber \\
&& 
 =    \sum_{x} O(x)   \sum_y O(y)
 \frac{2}{T} \int_0^{T} dt  \int_0^{T-t} d\tau  
\left[   \langle x \vert e^{w \tau } \vert y \rangle  - \langle x  \vert e^{w (\tau+t) } \vert x_0 \rangle \right] 
\langle y \vert e^{wt} \vert x_0 \rangle
\label{cumArho}
\end{eqnarray}
or equivalently in terms of
 the difference of Eq. \ref{defdelta},
\begin{eqnarray}
V_{ [{\cal O}_T] } &&
=  \sum_{x} O(x)   \sum_y O(y)
 \frac{2}{T}\int_0^{T} dt  \int_0^{T-t} d\tau  
\left[  \Delta_{\tau}(x,y)  - \Delta_{t+\tau}(x,x_0) \right] 
\left[ P_*(y) + \Delta_t(y,x_0) \right]
\label{cumArhobis}
\end{eqnarray}
As a consequence, the asymptotic rescaled variance is also governed by the Green function $G$
\begin{eqnarray}
V_{ [{\cal O}_T] } \opsimeq_{T \to +\infty}  
\sum_{x} O(x)   \sum_y O(y)
2   \left[ \int_0^{+\infty} d\tau \Delta_{\tau}(x,y)   \right]  P_*(y) 
=  \sum_{x} O(x)   \sum_y O(y) 2    G(x,y)  P_*(y) 
\label{varArho}
\end{eqnarray}
in agreement with the perturbative formula of Eq. \ref{energy2r} as it should.


\subsection{ Discussion }

In summary, the Green function $G$ 
plays a central role to characterize the typical convergence properties of empirical time-averaged
observables :

(i) it governs the convergence of the averaged value $< \rho_T(x) >_{Traj} $ of the empirical density $\rho_T(x) $
towards the steady state $P_*(x) $(Eq \ref{cvrhoav}),
and thus the convergence of the averaged value $< {\cal O}_T >_{Traj}  $ of any empirical time-averaged observable 
${\cal O}_T$ of the position towards its the steady value $O_*$ (Eq. \ref{avArho}).

(ii) it governs the scaled variance $V_{ [{\cal O}_T] } $ of any empirical time-averaged observable ${\cal O}_T$
of the position (Eq. \ref{cumArho}).

(iii) it governs the higher order cumulants via
the systematic perturbation theory recalled in subsection \ref{subsec_per}.

(iv) it also governs the cumulants of more general additive observables
involving the flows, as described for Markov chains either in discrete time or in continuous time 
in the context of Ruelle thermodynamic formalism for Markov trajectories
 \cite{c_ruelle}.


\subsection{ Diffusion processes in continuous space in dimension $d$ : adaptation to compute the Green function}

\label{subsec_fp}

When the Markov jump process of Eq. \ref{master} 
is replaced by the Fokker-Planck dynamics of Eq. \ref{fokkerplanck},
i.e. when the Markov matrix $w$ of Eq. \ref{master}
is replaced by the differential operator 
\begin{eqnarray}
 {\cal F}_{\vec x}    =  \partial_{\vec x} \left[ - \vec F(\vec x)  + D(\vec x)  \partial_{\vec x}  \right]
\label{generator}
\end{eqnarray}
most of the properties discussed above can be directly translated
by replacing the discrete sums over the discrete configuration $x$ by integrals over the
continuous position $\vec x$.

So here, we only mention that the explicit computation of the Green function
via the two methods described in subsection \ref{subsec_practice},
has to be adapted as follows :

(i) either one considers that the initial position $\vec x_0$ is a parameter, and the function $G(.,\vec x_0) $
can be computed via the forward equation analog to Eq \ref{Ginvright} 
\begin{eqnarray}
P_{*}(\vec x)   -\delta^d(\vec x - \vec x_0)  = {\cal F}_{\vec x}   G (\vec x, \vec x_0) =
 \partial_{\vec x} \left[ - \vec F(\vec x) G (\vec x, \vec x_0) + D(\vec x)  \partial_{\vec x} G (\vec x, \vec x_0) \right]
  \label{fpginverseright}
\end{eqnarray}
with the orthogonality condition analogous to Eq. \ref{Gorthogright} 
\begin{eqnarray}
0  = \int d^d \vec x G (\vec x, \vec x_0)
  \label{fporthoright}
\end{eqnarray}

(ii) or one considers instead that the final position $\vec x$ is a parameter, and the function $G(\vec x,.) $
can be computed by the backward equation analog to
Eq. \ref{Ginvleft} that involves the adjoint operator of Eq. \ref{generator}
\begin{eqnarray}
 {\cal F}^{\dagger}_{\vec x_0}     =   \left[  \vec F(\vec x_0)  +   \partial_{\vec x_0} \   D(\vec x_0)\right]  \partial_{\vec x_0}
\label{adjoint}
\end{eqnarray}
and that reads
\begin{eqnarray}
P_{*}(\vec x)   -\delta^d(\vec x - \vec x_0)  = {\cal F}^{\dagger}_{\vec x_0}  G (\vec x, \vec x_0)
=  \left[  \vec F(\vec x_0)  +   \partial_{\vec x_0} \   D(\vec x_0)\right]  \partial_{\vec x_0}G (\vec x, \vec x_0)
  \label{fpginverse}
\end{eqnarray}
with the orthogonality condition analogous to Eq. \ref{Gorthogleft}
\begin{eqnarray}
0  = \int d^d \vec x_0   G (\vec x, \vec x_0) P_{eq}(\vec x_0) 
= G (\vec x, \vec x) +  \int d^d \vec x_0   \left[ G (\vec x, \vec x_0) - G (\vec x, \vec x) \right] P_{eq}(\vec x_0) 
  \label{fportholeft}
\end{eqnarray}

Again, the method (ii) is technically simpler because Eq. \ref{fpginverse} is a first-order differential
equation for the partial derivative $\partial_{\vec x_0}G (\vec x, \vec x_0) $.
Again this can be also understood via the link with the MFPT $\tau(\vec x, \vec x_0)$ 
discussed in the subsection \ref{subsec_firstpassage}.


\section{ Explicit Green function for the Detailed-Balance Markov chain for $M$ }

\label{app_greenDBjump}

As explained in detail in Appendix \ref{sec_green},
the typical convergence properties of empirical observables are governed
by the Green function $G$.
For the generator of Eq. \ref{markovwDB},
the Green function $G(M,M_0)$ can be computed via the backward Eq. \ref{Ginvleft} with respect to the initial magnetization 
\begin{eqnarray}
P_{eq}(M) - \delta_{ M,M_0}  && = \sum_{M_0'}  G(M,M_0')  w(M_0',M_0)  
\nonumber \\
&& =  
\left[ G(M,M_0+2) - G(M,M_0) \right] W^+(M_0)
+   \left[ G(M,M_0-2) - G(M,M_0) \right] W^-(M_0)
 \label{Ginvleftm}
\end{eqnarray}
with the orthogonality condition of \ref{Gorthogleft}
 \begin{eqnarray}
0  =      \sum_{M_0=-N }^{N}    G(M,M_0) P_{eq}(M_0) 
 = G(M,M) +   \sum_{M_0 \ne M }  \left[ G(M,M_0) - G(M,M) \right] P_{eq}(M_0) 
\label{Gorthogleftm}
\end{eqnarray}

\subsection{ First step : computing the differences $g_M(M_0+1) \equiv G(M,M_0+2) - G(M,M_0) $ }

In Eq. \ref{Ginvleftm}, the final magnetization $M$ plays the role of a parameter,
so it is convenient to introduce the following simplified notation for the differences labelled by their middle-point $(M_0+1)$
 \begin{eqnarray}
g_M(M_0+1) \equiv G(M,M_0+2) - G(M,M_0)
\label{gdiff}
\end{eqnarray}
in order to rewrite Eq. \ref{Ginvleftm} as
\begin{eqnarray}
P_{eq}(M) - \delta_{ M,M_0}    = g_M(M_0+1)  W^+(M_0)- g_M(M_0-1) W^-(M_0)
 \label{recgdb}
\end{eqnarray}

(i) In the region $M_0 \in\{-N,-N+2,...,M-2\}$,
Eq. \ref{recgdb} corresponds to the recursion
\begin{eqnarray}
 g_M(M_0+1) =\frac{ g_M(M_0-1) W^-(M_0)+ P_{eq}(M) }{W^+(M_0)  }  
 \label{Ginvleftmsmaller}
\end{eqnarray}
and determines the boundary value for $M_0=-N$ where $W^-(N)=0$ 
\begin{eqnarray}
  g_M(-N+1) =\frac{ P_{eq}(M)}{W^+(-N)  }  
 \label{Ginvleftmsmallercln}
\end{eqnarray}
Using the Detailed-Balance condition \ref{mchainDB}, the solution of this recursion reads 
\begin{eqnarray}
g_M(M_0+1)  = \frac{ P_{eq}(M) }{W^+(M_0) P_{eq}(M_0)}
 \sum_{ y=-N }^{M_0}  P_{eq}(y)  
\ \ \ \    \ \ {\rm for } \  \in M_0 \in \{-N,...,M-2\}
  \label{gbelowsimpli}
\end{eqnarray}

(ii) In the region $M_0 \in \{M+2,...,N\}$, Eq. \ref{recgdb} corresponds to the recursion
\begin{eqnarray}
g_M(M_0-1)  = \frac{ g_M(M_0+1) W^+(M_0) - P_{eq}(M) }{W^-(M_0) }
 \label{Ginvleftmbig}
\end{eqnarray}
and determines the boundary value for $M_0=N$ where $W^+(N)=0$
\begin{eqnarray}
g_M(N-1)= - \frac{ P_{eq}(M) }{ W^-(N)}
 \label{Ginvleftmbigcln}
\end{eqnarray}
Using the Detailed-Balance condition \ref{mchainDB}, the solution of this recursion reads 
\begin{eqnarray}
g_M(M_0-1)  = - \frac{ P_{eq}(M) }{W^-(M_0) P_{eq}(M_0)}
 \sum_{ y=M_0}^{N}  P_{eq}(y)  
\ \ \ \ \   \ \ {\rm for } \ M_0 \in \{M+2,...,N\}
 \label{gabovesimpli}
\end{eqnarray}

(iii)  Using $g_M(M-1) $ obtained in Eq. \ref{gbelowsimpli} for $M_0=M-2$
\begin{eqnarray}
g_M(M-1)  = 
\frac{ P_{eq}(M) }{W^+(M-2) P_{eq}(M-2)}
 \sum_{ y=-N }^{M-2}  P_{eq}(y)
   \label{gsmallsolsimplilast}
\end{eqnarray}
and using $g_M(M+1)$ obtained Eq. \ref{gabovesimpli} for $M_0=M+2$
\begin{eqnarray}
g_M(M+1)  = 
- \frac{ P_{eq}(M) }{W^-(M+2) P_{eq}(M+2)}
 \sum_{ y=M+2 }^{N}  P_{eq}(y)  
 \label{gbigsolsimplilast}
\end{eqnarray}
one obtains that Eq. \ref{recgdb} for $M_0=M$ 
reads using again the Detailed-Balance condition of Eq. \ref{mchainDB}
\begin{eqnarray}
P_{eq}(M) - 1  &&  =  
g_M(M+1)  W^+(M)
- g_M(M-1)    W^-(M)
 =  - 
 \sum_{ y=M+2 }^{N}  P_{eq}(y)  
 -     
 \sum_{ y=-N }^{M-2}  P_{eq}(y)
 \label{gmiddle}
\end{eqnarray}
so that it is automatically satisfied as a consequence of the normalization of $P_{eq}(.)$.


\subsection{ Second step : computing the Green function $G(M,M_0) $ }

Now  that all the differences $ g_M(M_0+1)$ of Eq. \ref{gdiff} have been computed for each link,
one can obtain the differences $[G(M,M_0)-G(M,M) ]$ as follows. 

(i) For $M_0 \in \{-N,-N+2,...,M-2\}$, using the solution of Eq. \ref{gbelowsimpli},
one obtains 
\begin{eqnarray}
 G(M,M_0) - G(M,M) && =  \sum_{ x=M_0 }^{M-2}  \left[ G(M,x) - G(M,x+2) \right]
 = - \sum_{ x=M_0 }^{M-2} g_M(x+1)
 \nonumber \\
 && =  - P_{eq}(M)\sum_{x=M_0 }^{M-2} 
 \frac{ 1 }{W^+(x) P_{eq}(x)}
 \sum_{ y=-N }^{x}  P_{eq}(y)  
  \ \ \ {\rm for } \ \  M_0 \in \{-N,-N+2,...,M-2\}
 \label{GDBjumpbelow}
\end{eqnarray}

(ii) For $M_0 \in \{M+2,...,N\}$, using the solution of Eq. \ref{gabovesimpli},
one obtains 
\begin{eqnarray}
 G(M,M_0) - G(M,M) && =  \sum_{ x=M+2 }^{M_0}  \left[ G(M,x) - G(M,x-2)\right]
 = \sum_{x=M+2 }^{M_0} g_M(x-1)
 \nonumber \\
 && =  - P_{eq}(M)\sum_{x=M+2}^{M_0}
   \frac{ 1 }{W^-(x) P_{eq}(x)}
 \sum_{y=x}^{N}  P_{eq}(y)  
 \ \ \ {\rm for } \ \ M_0 \in \{M+2,...N-2,N\}
\label{GDBjumpabove}
\end{eqnarray}

(iii) Finally, the reference value $G(M,M)$ at coinciding point $M_0=M$
has to be determined from the orthogonality condition of Eq. \ref{Gorthogleftm}

 \begin{eqnarray}
 G(M,M) && = -    \sum_{M_0 \ne M }  \left[ G(M,M_0) - G(M,M) \right] P_{eq}(M_0) 
\nonumber \\
&&  =  -    \sum_{M_0 =-N }^{M-2} P_{eq}(M_0)  \left[ G(M,M_0) - G(M,M) \right] 
 -    \sum_{M_0 =M+2 }^{N} P_{eq}(M_0) \left[ G(M,M_0) - G(M,M) \right] 
\nonumber \\ &&   
 = P_{eq}(M) \sum_{ M_0=-N }^{M-2} 
 P_{eq}(M_0)
    \left[ \sum_{ x=M_0 }^{M-2} 
 \frac{ 1 }{W^+(x) P_{eq}(x)}
 \sum_{ y=-N }^{x}  P_{eq}(y)    \right] 
\nonumber \\
&&+ P_{eq}(M) \sum_{ M_0=M+2 }^{N} 
P_{eq}(M_0)
 \left[   \sum_{ x=M+2 }^{M_0}
   \frac{ 1 }{W^-(x) P_{eq}(x)}
 \sum_{ y=x }^{N}  P_{eq}(y) \right]
\label{GDBjumpcoinciding}
\end{eqnarray}


\subsection{ Link with the 
Mean-First-Passage-Time $\tau(M,M_0)$ at magnetization $M$ when starting at 
magnetization $M_0$ }

In order to see more clearly the physical meaning of the solution obtained above,
 it is convenient to use the rewriting of Eqs \ref{greenandfirst}
and \ref{Gorthoglefttau}
\begin{eqnarray}
G(M,M_0)  && = G(M,M)   - P_{eq}(M) \tau(M,M_0)  
\nonumber \\
 G(M,M) && = P_{eq}(M)   \sum_{M_0=-N  }^N \tau(M,M_0) P_{eq}(M_0)
\label{greentauDBM}
\end{eqnarray}
in terms of the 
Mean-First-Passage-Time $\tau(M,M_0)$ at magnetization $M$ when starting at 
magnetization $M_0$, with the two different expressions for $M_0<M$ and $M_0>M$
respectively
\begin{eqnarray}
 \tau(M,M_0) && =\sum_{x=M_0 }^{M-2} 
 \frac{ 1 }{W^+(x) P_{eq}(x)}
 \sum_{ y=-N }^{x}  P_{eq}(y)  
  \ \ \ {\rm for } \ \  M_0 \in \{-N,-N+2,...,M-2\}
\nonumber \\
 \tau(M,M_0) && = \sum_{x=M+2}^{M_0}
   \frac{ 1 }{W^-(x) P_{eq}(x)}
 \sum_{y=x}^{N}  P_{eq}(y)  
\ \  \ \ \ {\rm for } \ \ M_0 \in \{M+2,...N-2,N\}
\label{tauDBM}
\end{eqnarray}

For later purposes, let us mention the special case
of the MFPT $\tau^{tot}(M=0,M_0)$
to reach the magnetization $M=0$  when starting at $M_0 \ne 0$
\begin{eqnarray}
 \tau(M=0,M_0) 
&& =  \sum_{x=M_0 }^{-2} 
 \frac{ 1 }{W^+(x) P_{eq}(x)}
 \sum_{ y=-N }^{x}  P_{eq}(y)  
    \ \ {\rm for } \ \ M_0<0
\nonumber \\
 \tau(M=0,M_0) 
&& = \sum_{x=2}^{M_0}
   \frac{ 1 }{W^-(x) P_{eq}(x)}
 \sum_{y=x}^{N}  P_{eq}(y)  
      \ \ {\rm for } \ \ M_0>0
   \label{tauDBhighzero}
\end{eqnarray}

Finally, the equilibrium time $\tau_{eq}(M_0)$ of Eq. \ref{taueqx0indeo} does not depend on the initial magnetization $M_0$
and can be thus evaluated for the special case $M_0=N$
 \begin{eqnarray}
\tau_{eq} = \tau_{eq}(M_0=N) =  \sum_M P_{eq}(M) \tau(M,M_0=N) 
= \sum_{M=-N}^{N-2}  P_{eq}(M)  \sum_{x=M+2}^{N}   \frac{ 1 }{W^-(x) P_{eq}(x)} \sum_{y=x}^{N}  P_{eq}(y)  
\label{taueqDBM}
\end{eqnarray}
Its behavior for large $N$ will be discussed later in subsection \ref{subsec_taueqdiff}.


\section{ Explicit Green function for Skew-Detailed-Balance Markov chain for $M$ }

\label{app_greenSDBjump}

For the Markov matrix of Eq. \ref{wblock},
the Green function $G^{\sigma \sigma_0}(M,M_0)$ can be computed via 
the backward Eq. \ref{Ginvleft} 
\begin{eqnarray}
&& \frac{ P_{eq}(M)}{2} - \delta^{\sigma \sigma_0} \delta_{ M,M_0}  
 = \sum_{\sigma_0'} \sum_{M_0'}  G^{\sigma \sigma_0'}(M,M_0')  w^{\sigma_0' \sigma_0}(M_0',M_0)  
\nonumber \\
&& =  
  \delta^{\sigma_0,+}   \left[ G^{\sigma +}(M,M_0+2) - G^{\sigma +}(M,M_0) \right] W^+(M_0) 
 +  \delta^{\sigma_0,-}   \left[ G^{\sigma -}(M,M_0-2) - G^{\sigma -}(M,M_0)  \right] W^-(M_0) 
\nonumber \\
&&  +
\left[ G^{\sigma +}(M,M_0) - G^{\sigma -}(M,M_0)  \right]  
\left[ \delta^{\sigma_0,-}  \Gamma^-(M_0)-  \delta^{\sigma_0,+}   \Gamma^+(M_0) \right]
 \label{Ginvskew}
\end{eqnarray}
with the orthogonality condition of Eq. \ref{Gorthogleft}
 \begin{eqnarray}
0=   \sum_{\sigma_0= \pm 1}   \sum_{M_0}    G^{\sigma \sigma_0}(M,M_0) P_{eq}(M_0) 
\label{Gorthogleftskew}
\end{eqnarray}

To compare with the detailed-balance dynamics of the previous section, 
we are only interested into empirical observables of the magnetization $M$, 
whose typical convergence properties will be governed by the total Green function
after summing over the supplementary variables $\sigma=\pm 1 $ and $\sigma_0=\pm 1 $
 \begin{eqnarray}
G^{tot}(M,M_0) \equiv \sum_{\sigma=\pm 1}  \sum_{\sigma_0=\pm 1}  G^{\sigma \sigma_0}(M,M_0)  
\label{Gtot}
\end{eqnarray}


\subsection{ Equations for the intermediate Green function ${\cal G}^{\sigma_0}_M(M_0)  \equiv \sum_{\sigma=\pm 1}   G^{\sigma \sigma_0}(M,M_0)   $  }

Since the goal is to compute only the total Green function $G^{tot}$ of Eq. \ref{Gtot},
it is convenient to introduce the intermediate Green function after summing over the parameter $\sigma=\pm 1$
 \begin{eqnarray}
{\cal G}^{\sigma_0}_M(M_0)  \equiv \sum_{\sigma=\pm 1}   G^{\sigma \sigma_0}(M,M_0)  
\label{Ginter}
\end{eqnarray}
that satisfies the following closed system 
\begin{eqnarray}
 P_{eq}(M) -  \delta_{ M,M_0}  
&& =    \left[ {\cal G}^{ +}_M(M_0+2) - {\cal G}^{ +}_M(M_0) \right] W^+(M_0) 
   + \left[ {\cal G}^{ -}_M(M_0) - {\cal G}^{ +}_M(M_0)  \right]    \Gamma^+(M_0)
\nonumber \\
 P_{eq}(M) -  \delta_{ M,M_0}  
&& =    \left[ {\cal G}^{ -}_M(M_0-2) - {\cal G}^{ -}_M(M_0)  \right] W^-(M_0) 
  + \left[ {\cal G}^{ +}_M(M_0) - {\cal G}^{ -}_M(M_0)  \right]    \Gamma^-(M_0)
 \label{Ginvskew0}
\end{eqnarray}
while Eq. \ref{Gorthogleftskew} reads
 \begin{eqnarray}
0=   \sum_{\sigma_0= \pm 1}   \sum_{M_0}   {\cal G}^{\sigma_0}_M(M_0) P_{eq}(M_0) 
=  \sum_{M_0} G^{tot}(M,M_0)P_{eq}(M_0) 
\label{Gorthogleftskew0}
\end{eqnarray}


\subsection{ First step : computing the differences of the Green function ${\cal G}^{ \sigma_0}_M(M_0) $ for the links of the ladder }

It is convenient to introduce the two elementary differences for the links of the ladder
\begin{eqnarray}
g_M(M_0) && \equiv   {\cal G}^{ -}_M(M_0) - {\cal G}^{ +}_M(M_0) 
\nonumber \\
g_M(M_0+1) && \equiv   {\cal G}^{ +}_M(M_0+2)  - {\cal G}^{ -}_M(M_0) 
 \label{gskew}
\end{eqnarray}
satisfying
\begin{eqnarray}
g_M(M_0) + g_M(M_0+1) && =  {\cal G}^{ +}_M(M_0+2) - {\cal G}^{ +}_M(M_0) 
\nonumber \\
g_M(M_0-1) + g_M(M_0) && =  {\cal G}^{ -}_M(M_0)   -  {\cal G}^{ -}_M(M_0-2) 
 \label{gskew2}
\end{eqnarray}
Then Eqs \ref{Ginvskew0} can be rewritten as the recursions
\begin{eqnarray}
 P_{eq}(M) -  \delta_{ M,M_0}  
&& =     g_M(M_0+1) W^+(M_0) 
   + g_M(M_0) \left[ W^+(M_0) +  \Gamma^+(M_0) \right]
\nonumber \\
P_{eq}(M) -  \delta_{ M,M_0}  
&& =    - g_M(M_0-1)  W^-(M_0) 
 -  g_M(M_0) \left[  W^-(M_0)  +  \Gamma^-(M_0) \right]
 \label{recgskew}
\end{eqnarray}
Using $\Gamma^+(M) + W^+(M)  =  \Gamma^-(M) + W^-(M) $ of Eq. \ref{diffgammaout},
one obtains that the sum of these two equations 
allows to eliminate the function $g_M(M') $ for even argument $M'$
and gives the following closed recursion for the function $g_M(M'+1) $ for odd argument $(M'+1)$
\begin{eqnarray}
 2 P_{eq}(M) - 2 \delta_{ M,M_0}  
 =     g_M(M_0+1) W^+(M_0)     - g_M(M_0-1)  W^-(M_0) 
 \label{recgskewsum}
\end{eqnarray}


\subsubsection { Solution in the region $M_0<M$ }

 In the region $M_0 \in \{-N,...,M-2\}$, Eq. \ref{recgskewsum}
corresponds to the recursion
\begin{eqnarray}
g_M(M_0+1) = \frac{  2 P_{eq}(M)  +  g_M(M_0-1)  W^-(M_0) }  {  W^+(M_0)    }
 \label{gladderrecb}
\end{eqnarray}
and determines the boundary value for $M_0=-N$ where $W^-(N)=0$
\begin{eqnarray}
g_M(-N+1) = \frac{  2P_{eq}(M)   }  {  W^+(-N)    }
 \label{gladderrecbbord}
\end{eqnarray}
Using the Skew-Detailed-Balance condition $W^-(M+2) P_{eq}( M+2) - W^+(M) P_{eq}(M) $ of Eq. \ref{skewDBproj},
the solution reads
\begin{eqnarray}
g_M(M_0+1)  =  \frac{ 2 P_{eq}(M)}{W^+(M_0) P_{eq}(M_0) } \sum_{ y=-N }^{M_0} P_{eq}(y)
    \ \ \ {\rm for } \ \ M_0 \in \{-N,...,M-2\}
 \label{goddbelow}
\end{eqnarray}
Now the first Eq. \ref{recgskew}
can be used to compute
$g_M(.) $ for even argument in terms of the solution for odd argument of Eq. \ref{goddbelow}
\begin{eqnarray}
 g_M(M_0) 
&&  = \frac{  P_{eq}(M) -    g_M(M_0+1) W^+(M_0) }   { W^+(M_0) +  \Gamma^+(M_0) }
 \nonumber \\ &&     = \frac{ 2 P_{eq}(M)}    { W^+(M_0) +  \Gamma^+(M_0) }
    \left[ - \frac{1}{2}     -  \frac{ 1}{ P_{eq}(M_0) } \sum_{ y=-N }^{M_0-2} P_{eq}(y)    \right]
       \ \ \ {\rm for } \ \ M_0 \in \{-N,...,M-2\}
 \label{gevenbelow}
\end{eqnarray}
For later purposes, it is useful to mention the final result for the sum 
\begin{eqnarray}
g_M(M_0) + g_M(M_0+1) && 
=  \frac{  P_{eq}(M) +   g_M(M_0+1) \Gamma^+(M_0) }   { W^+(M_0) +  \Gamma^+(M_0) }
\nonumber \\
&& =
\frac{  P_{eq}(M) }   { W^+(M_0) +  \Gamma^+(M_0) }
\left[ 1 +   
 \frac{ 2  \Gamma^+(M_0)}{W^+(M_0) P_{eq}(M_0) } \sum_{ y=-N }^{M_0} P_{eq}(y)
\right]
 \label{gbelowsum}
\end{eqnarray}
for $ M_0 \in \{-N,...,M-2\}$.


\subsubsection { Solution in the region $M_0>M$ }

 In the region $M_0 \in \{M+2,...,N\}$, Eq. \ref{recgskewsum}
corresponds to the recursion
\begin{eqnarray}
  g_M(M_0-1) 
 =   \frac{  g_M(M_0+1) W^+(M_0)  -   2P_{eq}(M)  }{  W^-(M_0) }
 \label{gladderreca}
\end{eqnarray}
and determines the boundary value for $M_0=N$ where $W^+(N)=0$
\begin{eqnarray}
  g_M(N-1) 
 =  -  \frac{   2 P_{eq}(M)  }{  W^-(N) }
 \label{gladderrecabord}
\end{eqnarray}
Using the Skew-Detailed-Balance condition $W^-(M+2) P_{eq}( M+2) - W^+(M) P_{eq}(M) $ of Eq. \ref{skewDBproj},
the solution reads
\begin{eqnarray}
g_M(M_0-1)  = 
-   \frac{2 P_{eq}(M)}{W^-(M_0) P_{eq}(M_0)} \sum_{ y=M_0 }^{N} P_{eq}(y) 
    \ \ \ {\rm for } \ \ M_0 \in \{M+2,...,N\}
 \label{goddabove}
\end{eqnarray}
Now the second Eq. \ref{recgskew}
can be used to compute
$g_M(.) $ for even argument in terms of the solution for odd argument of Eq. \ref{goddabove}
\begin{eqnarray}
 g_M(M_0) 
&&  =  -  \frac{ P_{eq}(M)     + g_M(M_0-1)  W^-(M_0)  }{W^-(M_0)  +  \Gamma^-(M_0) }
\nonumber \\
&&   = \frac{2 P_{eq}(M)}{W^-(M_0)  +  \Gamma^-(M_0) }
   \left[  \frac{1}{2}   + \frac{1}{ P_{eq}(M_0)} \sum_{ y=M_0+2 }^{N} P_{eq}(y)    \right]
    \ \ \ {\rm for } \ \ M_0 \in \{M+2,...,N\}
 \label{gevenabove}
\end{eqnarray}

For later purposes, it is useful to mention the final result for the sum 
\begin{eqnarray}
 g_M(M_0-1) + g_M(M_0) 
 &&  =    \frac{ g_M(M_0-1)  \Gamma^-(M_0) - P_{eq}(M)       }{W^-(M_0)  +  \Gamma^-(M_0) }
 \nonumber \\
 &&
 = -   \frac{  P_{eq}(M)     }{W^-(M_0)  +  \Gamma^-(M_0) }
 \left[ 1   +  \frac{2  \Gamma^-(M_0) }{W^-(M_0) P_{eq}(M_0)} \sum_{ y=M_0 }^{N} P_{eq}(y)  \right]
 \label{gabovesum}
\end{eqnarray}
for $M_0 \in \{M+2,...,N\}$.


\subsubsection { Solution at coinciding points $M_0=M$ }

Eq. \ref{recgskewsum} for $M_0=M$
involves the below-solution of Eq \ref{goddbelow} for the last value $M_0=M-2$
  \begin{eqnarray}
g_M(M-1)  =  \frac{ 2 P_{eq}(M)}{W^+(M-2) P_{eq}(M-2) }
 \sum_{y=-N }^{M-2} P_{eq}(y)
 \label{goddbelowlast}
\end{eqnarray}
and the above-solution of Eq \ref{goddabove} for the last value $M_0=M+2$
\begin{eqnarray}
g_M(M+1)  = 
-   \frac{ 2 P_{eq}(M)}{W^-(M+2) P_{eq}(M+2)} \sum_{y=M+2 }^{N} P_{eq}(y) 
 \label{goddabovelast}
\end{eqnarray}
Using the Skew-Detailed-Balance condition of Eq. \ref{skewDBproj},
one obtains that Eq. \ref{recgskewsum} for $M_0=M$
\begin{eqnarray}
 P_{eq}(M) -  1  
=     \frac{ g_M(M+1) W^+(M)     - g_M(M-1)  W^-(M) }{2}
=  -   \sum_{ y=M+2}^{N} P_{eq}(y)    
    -   \sum_{ y=-N }^{M-2} P_{eq}(y)
 \label{recgskewsummiddle}
\end{eqnarray}
is automatically satisfied as a consequence of the normalization of $P_{eq}$.
The first Eq. \ref{recgskew} for $M_0=M$ that involves Eq. 
 \ref{goddabovelast} allows to compute the value of coinciding points $g_M(M_0=M)$
\begin{eqnarray}
  g_M(M) 
&&  = \frac{ P_{eq}(M) - 1  -    g_M(M+1) W^+(M)   }
{W^+(M) +  \Gamma^+(M)  }
    = \frac{1} {W^+(M) +  \Gamma^+(M)  } \left[ P_{eq}(M) - 1
+   2 \sum_{ y=M+2 }^{N} P_{eq}(y) 
 \right]
  \nonumber \\ &&
=  \frac{1} {W^+(M) +  \Gamma^+(M)  } \left[  \sum_{ y=M+2 }^{N} P_{eq}(y)   - \sum_{ y=-N }^{M-2} P_{eq}(y)   \right]
 \label{gevenmiddle}
\end{eqnarray}


\subsection{ Second step : computing the total Green function $G^{tot}(M,M_0) $ }

The goal is now to compute the total Green function $G^{tot}(M,M_0)  $ of Eq.\ref{Gtot}
\begin{eqnarray}
G^{tot}(M,M_0) = \sum_{\sigma_0=\pm} {\cal G}^{\sigma_0}_M(M_0) = 
 {\cal G}^{+}_M(M_0) +  {\cal G}^{-}_M(M_0) 
  \label{gtotcalG}
\end{eqnarray}

(i) In the region $M_0 \in \{-N,...,M-2\}$, one can use the differences of Eq. \ref{gskew2}
to compute
\begin{eqnarray}
G^{tot}(M,M_0) - G^{tot}(M,M)  
&& = \sum_{x=M_0}^{M-2} \left[ {\cal G}^{+}_M(x)  - {\cal G}^{+}_M(x+2)   \right]
 + \sum_{x=M_0}^{M-2} \left[ {\cal G}^{-}_M(x)  - {\cal G}^{-}_M(x+2)   \right]
\nonumber \\
&& = 
- \sum_{x=M_0}^{M-2} \left[ g_M(x) + g_M(x+1)  \right]
 -  \sum_{x=M_0}^{M-2} \left[ g_M(x+1) + g_M(x+2)   \right]
\nonumber \\
&& = 
-  2 \sum_{x=M_0}^{M-2} \left[ g_M(x) + g_M(x+1) \right]
+g_M(M_0) -g_M(M)
  \label{grefbelowpos}
\end{eqnarray}
in terms of the solution of Eqs \ref{gbelowsum} and \ref{gevenbelow} 
as well as Eq. \ref{gevenmiddle} for $g_M(M)$, leading to the final result given in
Eq. \ref{grefbelowposexpli}
of the text.


(ii) In the region $M_0 \in \{M+2,...,N\}$, one can use the differences of Eq. \ref{gskew2}
to obtain
\begin{eqnarray}
G^{tot}(M,M_0) - G^{tot}(M,M)  
&& = \sum_{x=M+2}^{M_0} \left[  {\cal G}^+_M(x)   -  {\cal G}^+_M(x-2)     \right]
+ \sum_{x=M+2}^{M_0} \left[  {\cal G}^-_M(x)   -  {\cal G}^-_M(x-2)     \right]
\nonumber \\
&& = 
 \sum_{x=M+2}^{M_0} \left[ g_M(x-2) + g_M(x-1)    \right]
+ \sum_{x=M+2}^{M_0} \left[  g_M(x-1) + g_M(x)    \right]
\nonumber \\
&& = 2 \sum_{x=M+2}^{M_0}   \left[  g_M(x-1) + g_M(x)    \right]
+ g_M(M) 
- g_M(M_0)
  \label{grefbelowneg}
\end{eqnarray}
in terms of the solution of Eqs \ref{gabovesum} and \ref{gevenabove} 
as well as Eq. \ref{gevenmiddle} for $g_M(M)$, leading to the final result given in 
Eq. \ref{grefbelownegfinal} of the text.

(iii) The value $G^{tot}(M,M)  $ at coinciding  points $M_0=M$ is fixed by
the orthogonality condition of \ref{Gorthogleftskew0}
that leads to the second Eq. \ref{greentauSDBM}
of the text.

\subsection { Mean-First-Passage-Time $\tau^{tot}(M,M_0)$ at magnetization $M$ when starting at 
magnetization $M_0$}

It is useful to write the solution found above
for $G^{tot}(M,M_0) $ in the same form as Eq. \ref{greentauDBM}
\begin{eqnarray}
G^{tot}(M,M_0)  && = G^{tot}(M,M)   - P_{eq}(M) \tau^{tot}(M,M_0)  
\nonumber \\
 G^{tot}(M,M) && = P_{eq}(M)   \sum_{M_0=-N  }^N \tau^{tot}(M,M_0) P_{eq}(M_0)
\label{greentauSDBM}
\end{eqnarray}
in terms of the MFPT $\tau^{tot}(M,M_0)$, with the two different expressions for $M_0 \in \{-N,...,M-2\}$
\begin{eqnarray}
&& \tau^{tot}(M,M_0) 
 = 
  2 \sum_{x=M_0}^{M-2} 
\frac{  1 }   { W^+(x) +  \Gamma^+(x) }
\left[ 1 +    \frac{ 2  \Gamma^+(x)}{W^+(x) P_{eq}(x) } \sum_{ y=-N }^{x} P_{eq}(y) \right]
 \label{grefbelowposexpli} \\
&&
+ \frac{ 1 }    { W^+(M_0) +  \Gamma^+(M_0) }
    \left[  1    +  \frac{ 2}{ P_{eq}(M_0) } \sum_{ y=-N }^{M_0-2} P_{eq}(y)    \right]
+  \frac{1} {\left[ W^+(M) +  \Gamma^+(M)\right] P_{eq}(M)  } \left[  \sum_{ y=M+2 }^{N} P_{eq}(y)   - \sum_{ y=-N }^{M-2} P_{eq}(y)   \right]
 \nonumber 
\end{eqnarray}
and for $M_0  \in \{M+2,...,N\}$ respectively
\begin{eqnarray}
&& \tau^{tot}(M,M_0) 
 =  2 \sum_{x=M+2}^{M_0}   
   \frac{  1     }{W^-(x)  +  \Gamma^-(x) }
 \left[ 1   +  \frac{2  \Gamma^-(x) }{W^-(x) P_{eq}(x)} \sum_{ y=x }^{N} P_{eq}(y)  \right]
   \label{grefbelownegfinal}
 \\
&&
+  \frac{ 1}{W^-(M_0)  +  \Gamma^-(M_0) }
   \left[   1  + \frac{2}{ P_{eq}(M_0)} \sum_{ y=M_0+2 }^{N} P_{eq}(y)    \right]
+  \frac{1} {\left[ W^-(M) +  \Gamma^-(M)\right] P_{eq}(M)  } 
\left[  \sum_{ y=-N }^{M-2} P_{eq}(y)   - \sum_{ y=M+2 }^{N} P_{eq}(y)    \right]
\nonumber
\end{eqnarray}
that should be compared to their detailed-balance counterpart of Eqs \ref{tauDBM}.
However, the physical meaning of these formula
is somewhat obscured by the presence of various contributions.
In addition, one should take into account the regions where the switching rates $\Gamma^+(M) $
or $\Gamma^-(M) $ vanish according to Eq. \ref{gammathetaV}.
So let us now focus on the following special case.
In the high-temperature phase or the critical point $\beta J \leq \beta_c J=1$,
where the switching rates are given by Eq. \ref{gammahigh},
the MFPT $\tau^{tot}(M=0,M_0)$
to reach the magnetization $M=0$ when starting at $M_0 \ne 0$ simplifies into
\begin{eqnarray}
 \tau^{tot}(M=0,M_0) 
&& =   2 \sum_{x=M_0}^{-2}  \frac{  1 }   { W^+(x)  }
+ \frac{ 1 }    { W^+(M_0)  }
    \left[  1    +  \frac{ 2}{ P_{eq}(M_0) } \sum_{ y=-N }^{M_0-2} P_{eq}(y)    \right]
    \ \ {\rm for } \ \ M_0<0
\nonumber \\
 \tau^{tot}(M=0,M_0) 
&& =  2 \sum_{x=2}^{M_0}      \frac{  1     }{W^-(x)   }
+  \frac{ 1}{W^-(M_0) }
   \left[   1  + \frac{2}{ P_{eq}(M_0)} \sum_{ y=M_0+2 }^{N} P_{eq}(y)    \right]
      \ \ {\rm for } \ \ M_0>0
   \label{tauSDBhighzero}
\end{eqnarray}
The first terms involving sums of the inverses of the rates $\frac{1}{W^{\pm}(x)}$ towards zero
clearly show the directed character of the motion in contrast to the Detailed-Balance counterpart
of Eq. \ref{tauDBhighzero}.


\section{ Explicit Green function for the detailed-balance diffusion process for $m$}

\label{app_greenDBdiff}

The Green function $G(m,m_0)$ 
can be found by solving the backward Eq. \ref{fpginverse} 
\begin{eqnarray}
p_{eq}(m)   -\delta(m - m_0)  
&& =  \left[  F(m)  +   \partial_{m_0} \   D \right]  \partial_{m_0}G (m, m_0)
\nonumber \\
&& = D \left[  -  N V'(m) +   \partial_{m_0}  \right]  \partial_{m_0}G (m, m_0)
  \label{fpginversem}
\end{eqnarray}
with the orthogonality condition of Eq. \ref{fportholeft}
\begin{eqnarray}
0 = G (m,m) +  \int_{-1}^1 dm_0   \left[ G (m, m_0) - G (m, m) \right] p_{eq}(m_0) 
  \label{fportholeftm}
\end{eqnarray}

\subsection { First step : computing the derivative $g_m(m_0) \equiv \partial_{m_0} G(m,m_0)  $ }

Since $m$ plays the role of a parameter in Eq. \ref{fpginversem},
it is convenient to introduce the following notation 
\begin{eqnarray}
 g_m(m_0) \equiv   \partial_{m_0} G(m,m_0)  
   \label{gdiffm}
\end{eqnarray}
so that Eq. \ref{fpginversem} becomes the first order differential equation
\begin{eqnarray}
   \left[  -  N V'(m) +   \partial_{m_0}  \right]  g_m(m_0)=   \frac{ p_{eq}(m)   -\delta(m-m_0) }{D}
  \label{gfirst}
\end{eqnarray}
In the region $m_0 \in ]-1,m[$, the solution vanishing at the boundary $m_0 \to -1$ reads
\begin{eqnarray}
 g_m(m_0)= \frac{p_{eq}(m) }{D} e^{NV (m_0)}  \int_{-1}^{m_0}    dx   e^{- NV(x) } 
 = \frac{ p_{eq}(m) }{ D p_{eq}(m_0) } \int_{-1}^{m_0}    dx   p_{eq}(x)
 \ \ {\rm for } \ \ m_0 \in ]-1,m[
  \label{gfirstbelow}
\end{eqnarray}
In the region $m_0 \in ]m,1[$, the solution vanishing at the boundary $m_0 \to +1$ reads
\begin{eqnarray}
 g_m(m_0)= - \frac{p_{eq}(m) }{D} e^{NV (m_0)}  \int_{m_0}^1    dx   e^{- NV(x) } 
 = - \frac{ p_{eq}(m) }{ D p_{eq}(m_0) } \int_{m_0}^1    dx   p_{eq}(x)
 \ \ {\rm for } \ \ m_0 \in ]m,1[
  \label{gfirstabove}
\end{eqnarray}
The integration of Eq. \ref{gfirst} between $m_0=m-\epsilon$ and $m_0=m+\epsilon$ 
involves the discontinuity between the two solutions found above
\begin{eqnarray}
 - \frac{1}{D} =   g_m(m+\epsilon) -  g_m(m-\epsilon)
  = - \frac{ 1 }{ D  } \int_{m}^1    dx   p_{eq}(x)
  -  \frac{ 1 }{ D  } \int_{-1}^{m}    dx   p_{eq}(x)
  =  -  \frac{ 1 }{ D  } \int_{-1}^{1}    dx   p_{eq}(x)
  \label{gfirstd}
\end{eqnarray}
and is satisfied as a consequence of the normalization of $p_{eq}(m)$.

\subsection { Second step : computing the Green function $ G(m,m_0) $ }

The integration of Eq. \ref{gdiff} 
allows to compute the differences
\begin{eqnarray}
G(m,m_0)  - G(m,m) = \int_m^{m_0} dy g_m(y)
   \label{gdiffinteg}
\end{eqnarray}
In the region $m_0 \in ]-1,m[$, the solution \ref{gfirstbelow} yields the difference
\begin{eqnarray}
G(m,m_0)  - G(m,m) = -  \int_{m_0}^m dy g_m(y) 
= - \frac{ P_{eq}(m)}{D}  \int_{m_0}^m    \frac{ dy }{ P_{eq}(y) } \int_{-1}^{y}    dx   P_{eq}(x)
\ \ {\rm for } \ \ m_0 \in ]-1,m[
  \label{gsecondbelow}
\end{eqnarray}
In the region $m_0 \in ]m,1[$, the solution \ref{gfirstabove} yields the difference
\begin{eqnarray}
G(m,m_0)  - G(m,m) = \int_m^{m_0} dy g_m(y) 
= - \frac{ P_{eq}(m)}{D}  \int_m^{m_0}    \frac{ dy }{ P_{eq}(y) } \int_{y}^{1}    dx   P_{eq}(x)
 \ \ {\rm for } \ \ m_0 \in ]m,1[
  \label{gsecondabove}
\end{eqnarray}
The value $G(m,m)$ at coinciding point $m_0=m$ is determined by the orthogonality condition of Eq. \ref{fportholeftm}
\begin{eqnarray}
&& G (m,m)  = -  \int_{-1}^m dm_0   \left[ G (m, m_0) - G (m, m) \right] p_{eq}(m_0) 
 - \int_{m}^1 dm_0   \left[ G (m, m_0) - G (m, m) \right] p_{eq}(m_0) 
 \nonumber \\
&& =  \frac{ P_{eq}(m)}{D}   \int_{-1}^m dm_0    p_{eq}(m_0) 
   \int_{m_0}^m    \frac{ dy }{ P_{eq}(y) } \int_{-1}^{y}    dx   P_{eq}(x)
 + \frac{ P_{eq}(m)}{D}  \int_{m}^1 dm_0    p_{eq}(m_0) 
    \int_m^{m_0}    \frac{ dy }{ P_{eq}(y) } \int_{y}^{1}    dx   P_{eq}(x)
  \label{fportholeftmfinal}
\end{eqnarray}

\subsection{ Link with the 
Mean-First-Passage-Time $\tau(m,m_0)$ at magnetization $m$ when starting at 
magnetization $m_0$ }

To see more clearly the physical meaning of the solution obtained above,
 it is convenient to use again the rewriting of Eqs \ref{greenandfirst}
and \ref{Gorthoglefttau}
\begin{eqnarray}
G(m,m_0)  && = G(m,m)   - p_{eq}(m) \tau(m,m_0)  
\nonumber \\
 G(m,m) && = p_{eq}(m)   \int_{-1}^{+1} dm_0 \tau(m,m_0) p_{eq}(m_0)
\label{greentauDBm}
\end{eqnarray}
in terms of the 
Mean-First-Passage-Time $\tau(m,m_0)$ at magnetization $m$ when starting at 
magnetization $m_0$, with the two different expressions for $m_0<m$ and $m_0>m$
respectively, using the diffusion coefficient $D=\frac{2}{N}$  of Eq. \ref{dcoef}
and the equilibrium distribution of Eq. \ref{Peqm}
\begin{eqnarray}
\tau(m,m_0) && =  \int_{m_0}^m    \frac{ dy }{ D p_{eq}(y) } \int_{-1}^{y}    dx   p_{eq}(x)
=
 \frac{N}{2} \int_{m_0}^m   dy \int_{-1}^{y}    dx   e^{ N \left[ V(y)-V(x) \right] }
\ \ {\rm for } \ \ m_0 \in ]-1,m[
 \nonumber \\
\tau(m,m_0) && =  
 \int_m^{m_0}    \frac{ dy }{ D p_{eq}(y) } \int_{y}^{1}    dx   p_{eq}(x)
=  \frac{N}{2} \int_m^{m_0}   dy\int_{y}^{1}    dx   e^{ N \left[ V(y)-V(x) \right] }
 \ \ {\rm for } \ \ m_0 \in ]m,1[
  \label{tauDBdiff}
\end{eqnarray}


\subsection{ Equilibrium time $\tau_{eq} $ }

\label{subsec_taueqdiff}

The equilibrium time $\tau_{eq}(m_0)$ of Eq. \ref{taueqx0indeo} does not depend on the initial magnetization $m_0$
and can be thus evaluated for the special case $m_0=1$
 \begin{eqnarray}
\tau_{eq} && = \tau_{eq}(m_0=1) =  \int_{-1}^{+1} dm p_{eq}(m) \tau(m,m_0=1) 
 \nonumber \\ &&
 =  \frac{N}{2 \int_{-1}^{+1} dz e^{-  N  V(z) }} \int_{-1}^{+1} dm   e^{-  N  V(m) } 
  \int_m^{1}   dy\int_{y}^{1}    dx   e^{ N \left[ V(y)-V(x) \right] }
\label{taueqDBdiff}
\end{eqnarray}

For large $N$, in the high-temperature phase and at criticality, it is the expansion around $m=0$
of Eq. \ref{vmseries} that will be important, with its leading contribution
\begin{eqnarray}
 V(m) && = (1  - \beta J ) \frac{m^2 }{2}  +O(m^4)  \ \ \ \ \ {\rm for } \ \ \beta J< \beta_c J=1
 \nonumber \\
 V(m) && =  \frac{m^4}{12} +O(m^6)\ \ \ \ \ \ \ \ \ \ \ \ \ {\rm for } \ \ \beta=\beta_c
\label{vmserieshc}
\end{eqnarray}
In order to analyze both cases together, let us assume that the leading behavior is of order $p$
\begin{eqnarray}
 V(m) = \alpha m^p + o(m^p)
\label{vmp}
\end{eqnarray}
and let us rescale accordingly all the integration variables as $ \tilde m= (\alpha N)^{\frac{1}{p} } m$ in Eq. \ref{taueqDBdiff}
to obtain
 \begin{eqnarray}
\tau_{eq} \opsimeq_{N \to +\infty} \frac{N}{(\alpha N)^{\frac{2}{p} } } 
 \frac{1 }{2 \int_{-\infty}^{+\infty} d{\tilde z} e^{-     {\tilde z} ^p }} 
 \int_{-\infty}^{+\infty} d{\tilde m} e^{-     {\tilde m} ^p }
  \int_{\tilde m}^{+\infty}   d {\tilde y} \int_{\tilde y}^{+\infty}    d{\tilde x}   e^{ {\tilde y} ^p - {\tilde x} ^p  }
\label{taueqDBdiffrescal}
\end{eqnarray}
In the high-temperature phase with $p=2$ and $\alpha= (1  - \beta J )= (\beta_c-\beta) J$,
the equilibrium time does not depend on $N$ but diverges when one approaches the transition $\beta \to \beta_c$
 \begin{eqnarray}
\tau_{eq} \propto \frac{1}{\beta_c-\beta  } \ \ \ \ \ {\rm for } \ \ \beta < \beta_c 
\label{taueqDBdiffrescalhigh}
\end{eqnarray}
in agreement with the previous study \cite{CurieW_slowingdown} based on the same choice of rates in Eq. \ref{DBchoice}.
Note however that in the discrete-time formulation where one counts instead the number of Markov-chain steps
\cite{TCV,weigel,vuceljaa,SDBcurierun}, the convergence-time is found to scale as $N$ in the high-temperature phase.

At criticality with $p=4$, the equilibrium time diverges with the system size $N$ as
 \begin{eqnarray}
\tau_{eq} \opsimeq_{N \to +\infty} \sqrt{N} \ \ \ \ \ {\rm at } \ \ \beta= \beta_c 
\label{taueqDBdiffrescalcriti}
\end{eqnarray}
in agreement with the general phenomenon of critical slowing down.
Again in the discrete-time formulation where one counts instead the number of Markov-chain steps
\cite{TCV,weigel,vuceljaa,SDBcurierun}, the convergence-time is found to scale as $N^{\frac{3}{2}}$ at criticality.

In the low-temperature phase $\beta>\beta_c$ with the shape with two minima for the potential $V(m)$ (Eq. \ref{vprime}),
the equilibrium time will instead diverge exponentially in $N$ as a consequence of the barrier $V(m=0)$ at the local maximum $m=0$ between the two minima $V_{min}=V(\pm m_{sp}(\beta) ) $ at $\pm m_{sp}(\beta) $
 \begin{eqnarray}
\tau_{eq} \oppropto_{N \to +\infty} e^ { N \left[ V(0) -V(m_{sp}(\beta) )  \right] }  \ \ \ \ \ {\rm at } \ \ \beta> \beta_c 
\label{taueqDBdifflow}
\end{eqnarray}


\section{ Explicit Green function for the run-and-tumble process in $m$ }

\label{app_greenSDBrun}

The dynamics of Eq. \ref{runmsmall} can be rewritten
\begin{eqnarray}
\partial_t
\begin{pmatrix} 
p_{t}^+( m )
 \\  p_{t}^-( m )
  \end{pmatrix}
= \begin{pmatrix} 
 -\partial_m \left[   p_{t}^+( m )  \right] 
  -   \gamma^+ (m) p_{t}^+( m ) 
  +  \gamma^- (m) p_{t}^-( m ) \\
 \partial_m \left[  p_{t}^-( m ) \right] 
       +   \gamma^+ (m) p_{t}^+( m ) 
  -  \gamma^- (m) p_{t}^-( m )
 \end{pmatrix} 
 \equiv 
  {\cal F}_{m} \begin{pmatrix} 
p_{t}^+( m )
 \\  p_{t}^-( m )
  \end{pmatrix}
\label{langevinmatrix}
\end{eqnarray}
with the generator
\begin{eqnarray}
  {\cal F}_{m} =  
  \begin{pmatrix} 
-   \gamma^+ (m) -  \partial_m       & \gamma^- (m)  \\
     \gamma^+ (m)   &    -  \gamma^- (m) +  \partial_m        
   \end{pmatrix} 
\label{fmatrix}
\end{eqnarray}
so that its adjoint reads
\begin{eqnarray}
  {\cal F}^{\dagger}_{m_0} =  
  \begin{pmatrix} 
-   \gamma^+ (m_0) +    \partial_{m_0}  & \gamma^+ (m_0)  \\
     \gamma^- (m_0)   &    -  \gamma^- (m_0)        -   \partial_{m_0}
   \end{pmatrix} 
\label{fmatrixdagger}
\end{eqnarray}
In the following, we will need the difference and the sum of the switching rates
of Eq. \ref{smallgamma}
that simply involve the derivative $V'(m)$ and its absolute value $\vert V' (m) \vert $
 \begin{eqnarray}
\gamma^+ (m) - \gamma^- (m) && =  N  V' (m) 
 \nonumber \\
\gamma^+ (m) +\gamma^- (m) && =  N  \vert V' (m) \vert
\label{smallgammasd}
\end{eqnarray}


\subsection{ Equations satisfied by the Green function $G^{\sigma \sigma_0}(m,m_0)$ }

As in the other cases discussed previously, the Green function $G^{\sigma \sigma_0}(m,m_0) $ can be computed via two methods :

(i) either one considers that $(m_0,\sigma_0)$ are parameters,
and the Green $G^{\sigma \sigma_0}(m,m_0)$ satisfies the forward system
\begin{eqnarray}
&&  \begin{pmatrix} 
\frac{ p_{eq}(m)}{2}   -\delta(m-m_0)      & \frac{ p_{eq}(m)}{2}  \\
\frac{ p_{eq}(m)}{2}    &   \frac{ p_{eq}(m)}{2}    -\delta(m-m_0)        
   \end{pmatrix}
    = 
    {\cal F}_{m} 
      \begin{pmatrix} 
  G^{++}(m,m_0)  & G^{+-}(m,m_0)  \\
  G^{-+}(m,m_0)   &   G^{--}(m,m_0)      
   \end{pmatrix}   
  \label{eqgreenfp}
\end{eqnarray}
i.e. more explicitly for $\sigma=\pm$ 
\begin{eqnarray}
\frac{p_{eq}(m)}{2}   -\delta(m-m_0) \delta_{\sigma_0,+} &&  =  
-  \partial_m  G^{+\sigma_0}(m,m_0)
-   \gamma^+ (m)G^{+\sigma_0}(m,m_0) 
 + \gamma^- (m) G^{-\sigma_0}(m,m_0) 
\nonumber \\
\frac{p_{eq}(m)}{2}   -\delta(m-m_0) \delta_{\sigma_0,-} && = 
  \partial_m  G^{-\sigma_0}(m,m_0)   
+  \gamma^+ (m)  G^{+\sigma_0}(m,m_0)     -  \gamma^- (m)G^{-\sigma_0}(m,m_0)     
  \label{eqgreenfpexpli}
\end{eqnarray}
with the orthogonality condition
\begin{eqnarray}
0  = \sum_{\sigma=\pm 1} \int_{-1}^1 dm   G^{\sigma \sigma_0}(m,m_0)  
=  \int_{-1}^1 dm  \left[ G^{+ \sigma_0}(m,m_0)  +G^{- \sigma_0}(m,m_0) \right]
  \label{eqgreenfportho}
\end{eqnarray}

(ii) or one considers that $(m,\sigma)$ are parameters,
and the Green $G^{\sigma \sigma_0}(m,m_0)$ satisfies the backward system
\begin{eqnarray}
&&  \begin{pmatrix} 
\frac{ p_{eq}(m)}{2}   -\delta(m-m_0)      & \frac{ p_{eq}(m)}{2}  \\
\frac{ p_{eq}(m)}{2}    &   \frac{ p_{eq}(m)}{2}    -\delta(m-m_0)        
   \end{pmatrix}
    = 
    {\cal F}^{\dagger}_{m_0} 
      \begin{pmatrix} 
  G^{++}(m,m_0)  & G^{-+}(m,m_0)  \\
  G^{+-}(m,m_0)   &   G^{--}(m,m_0)      
   \end{pmatrix}   
  \label{eqgreenleft0}
\end{eqnarray}
i.e. more explicitly for $\sigma=\pm$ 
\begin{eqnarray}
\frac{p_{eq}(m)}{2}  -\delta(m-m_0) \delta_{\sigma,+}
&& =     \partial_{m_0} G^{\sigma+}(m,m_0) + \gamma^+ (m_0) \left[  G^{\sigma-}(m,m_0) - G^{\sigma+}(m,m_0)\right]
\nonumber \\
\frac{p_{eq}(m)}{2}  -\delta(m-m_0) \delta_{\sigma,-}
&& =  -     \partial_{m_0}G^{\sigma-}(m,m_0)
+  \gamma^- (m_0) \left[ G^{\sigma+}(m,m_0)       -   G^{\sigma-}(m,m_0)  \right]         
  \label{rungleft}
\end{eqnarray}
with the orthogonality condition
\begin{eqnarray}
0  = \sum_{\sigma_0= \pm 1} \int_{-1}^1 dm_0   G^{\sigma \sigma_0}(m,m_0)  p_{eq}(m_0) 
=  \int dm_0  \left[ G^{\sigma +}(m,m_0)  
+G^{\sigma -}(m,m_0)   \right]  p_{eq}(m_0)
  \label{runortho}
\end{eqnarray}
As usual, the method (ii) is technically simpler. 


\subsection{ Equations for the intermediate Green function ${\cal G}^{\sigma_0}_m(m_0)  \equiv \sum_{\sigma=\pm 1}   G^{\sigma \sigma_0}(m,m_0)   $  }

Since we are only interested into 
empirical observables of the intensive magnetization $m$, 
 we only need to compute the total Green function
 \begin{eqnarray}
G^{tot}(m,m_0) \equiv \sum_{\sigma=\pm 1}  \sum_{\sigma_0=\pm 1}  G^{\sigma \sigma_0}(m,m_0)  
\label{Gtotm}
\end{eqnarray}
It is thus convenient to introduce
the intermediate Green function
 \begin{eqnarray}
{\cal G}^{\sigma_0}_m(m_0)  \equiv \sum_{\sigma=\pm 1}   G^{\sigma \sigma_0}(m,m_0)  
\label{Ginterm}
\end{eqnarray}
that satisfies closed equations after summing 
Eq. \ref{rungleft} over $\sigma=\pm$,
i.e. the two functions ${\cal G}^{\sigma_0=\pm}_m(m_0)$ satisfy the system 
\begin{eqnarray}
p_{eq}(m)  -\delta(m-m_0) 
&& =     \partial_{m_0} {\cal G}^{+}_m(m_0) + \gamma^+ (m_0) \left[  {\cal G}^{-}_m(m_0) - {\cal G}^{+}_m(m_0)\right]
\nonumber \\
p_{eq}(m)  -\delta(m-m_0) 
&& =  -     \partial_{m_0} {\cal G}^{-}_m(m_0)
+  \gamma^- (m_0) \left[ {\cal G}^{+}_m(m_0)      -  {\cal G}^{-}_m(m_0)  \right]         
  \label{rungleftsum}
\end{eqnarray}
It is then useful to introduce the difference
\begin{eqnarray}
g_m(m_0)   \equiv   {\cal G}^{-}_m(m_0) - {\cal G}^{+}_m(m_0)
  \label{differencegrun}
\end{eqnarray}
while the sum corresponds to the total Green function of Eq. \ref{Gtotm} that we wish to compute
\begin{eqnarray}
G^{tot}(m,m_0) \equiv  {\cal G}^{-}_m(m_0) + {\cal G}^{+}_m(m_0)
\label{Gtotsum}
\end{eqnarray}
The sum of the two Eqs \ref{rungleftsum} then gives a closed equation for $g_m(m_0) $
that involves the difference of the two switching rates $\gamma^+ (m_0) - \gamma^- (m_0)  =  N  V' (m_0)  $ 
of Eq. \ref{smallgammasd}
\begin{eqnarray}
2 p_{eq}(m)  - 2 \delta(m-m_0) 
&& =     - \partial_{m_0} g_m(m_0) + \left[ \gamma^+ (m_0) -  \gamma^- (m_0)\right]g_m(m_0) 
\nonumber \\
&& =     - \partial_{m_0} g_m(m_0) + N  V' (m_0)  g_m(m_0) 
  \label{closedsmallg}
\end{eqnarray}
that will be solved in the next subsection \ref{subsec_smallgsol}.

The difference of the two Eqs \ref{rungleftsum} involves the sum
of the two switching rates $\gamma^+ (m_0) + \gamma^- (m_0)  =  N  \vert V' (m_0) \vert $ 
of Eq. \ref{smallgammasd}
\begin{eqnarray}
0 && =     \partial_{m_0} G^{tot}(m,m_0) + \left[ \gamma^+ (m_0) +  \gamma^- (m_0)\right]g_m(m_0)
\nonumber \\
&& =     \partial_{m_0} G^{tot}(m,m_0) +  N  \vert V' (m_0) \vert  g_m(m_0)
  \label{eqgG}
\end{eqnarray}
while the sum of Eq. \ref{runortho} over $\sigma=\pm$ yields the following orthogonality 
condition for $G^{tot}(m,m_0) $
\begin{eqnarray}
0  =  \int dm_0  G^{tot}(m,m_0)  p_{eq}(m_0)
  \label{runorthosum}
\end{eqnarray}


\subsection{ Computing the difference $g_m(m_0) $ }

\label{subsec_smallgsol}

The closed Eq. \ref{closedsmallg} for $g_m(m_0)  $ can be solved as follows.

 In the region $m_0 \in ]-1,m[$, the solution that vanishes at the boundary $m_0 \to -1$ reads
\begin{eqnarray}
 g_m(m_0) && = - 2 p_{eq}(m)  e^{ NV(m_0) }    \int_{-1}^{m_0} dx   e^{ -NV(y) }  
 = - 2 \frac{p_{eq}(m)}{ p_{eq}(m_0)}  \int_{-1}^{m_0} dx p_{eq}(x)  
 \ \ {\rm for } \ \ m_0 \in ]-1,m[
  \label{grunbelow}
\end{eqnarray}

 In the region $m_0 \in ]m,1[$, the solution that vanishes at the boundary $m_0 \to 1$ reads
\begin{eqnarray}
 g_m(m_0) && = 2 p_{eq}(m)e^{ U(m_0) }    \int_{m_0}^{1} dx   e^{ -U(y) }  
  =  2 \frac{p_{eq}(m)}{ p_{eq}(m_0)}  \int_{m_0}^{1} dx p_{eq}(x)  
   \ \ {\rm for } \ \ m_0 \in ]m,1[
  \label{grunabove}
\end{eqnarray}

The integration of Eq. \ref{closedsmallg} between $m_0=m-\epsilon$ and $m_0=m+\epsilon$ corresponds to the discontinuity
 \begin{eqnarray}
  - 2  =   g_m(m-\epsilon) -g_m(m+\epsilon)
  = - 2  \int_{-1}^{m} dx p_{eq}(x)  - 2   \int_{m}^{1} dx p_{eq}(x)  = - 2   \int_{-1}^{1} dx p_{eq}(x)
  \label{closedsmallgdisc}
\end{eqnarray}
and is satisfied as a consequence of the normalization of $p_{eq}(m)$.


\subsection{ Computing the total Green function $G^{tot}(m,m_0) $ }

In order to compute the differences $G^{tot}(m,m_0) - G^{tot}(m,m) $, one needs to integrate 
Eq. \ref{eqgG}  
\begin{eqnarray}
     \partial_{m_0} G^{tot}(m,m_0) = -  N  \vert V' (m_0) \vert  g_m(m_0)
  \label{eqgGbis}
\end{eqnarray}
as follows.

In the region $m_0 \in ]-1,m[$, the solution of Eq. \ref{grunbelow} for $g_m(m_0) $ 
allows to compute  
\begin{eqnarray}
G^{tot}(m,m_0)  -  G^{tot}(m,m)
  =   \int_{m_0}^{m} dy N  \vert V' (y) \vert  g_m(y)    
  \label{fbelow}
\end{eqnarray}
In the region $m_0 \in ]m,1[$, the solution of Eq. \ref{grunabove} for $g_m(m_0) $
allows to compute
\begin{eqnarray}
G^{tot}(m,m_0)  -  G^{tot}(m,m)
  =  - \int_{m}^{m_0} dy N  \vert V' (y) \vert g_m(y)        
    \label{fabove}
\end{eqnarray}
Finally, the orthogonality condition of Eq. \ref{runorthosum} 
allows to compute the total Green function $G^{tot}(m,m_0) $
at coinciding points $m_0=m$ via
\begin{eqnarray}
 G^{tot} (m,m) && = -  \int_{-1}^m dm_0   \left[ G^{tot}(m,m_0)  - G(m,m)  \right] p_{eq}(m_0)
+   \int_{m}^1 dm_0  \left[ G^{tot}(m,m_0)  - G(m,m)  \right] p_{eq}(m_0) 
  \label{gtotrun}
\end{eqnarray}
and one obtains the final results summarized by Eqs \ref{greentauSDBm} and \ref{tauSDB}
of the text.

\subsection{  
Mean-First-Passage-Time $\tau^{tot}(m,m_0)$ at magnetization $m$ when starting at 
magnetization $m_0$ }

The solution found above can be rewritten as Eq. \ref{greentauDBm}
\begin{eqnarray}
G^{tot}(m,m_0)  && = G^{tot}(m,m)   - p_{eq}(m) \tau^{tot}(m,m_0)  
\nonumber \\
 G^{tot}(m,m) && = p_{eq}(m)   \int_{-1}^{+1} dm_0 \ \tau^{tot}(m,m_0) p_{eq}(m_0)
\label{greentauSDBm}
\end{eqnarray}
where the MFPT $\tau^{tot}(m,m_0) $
\begin{eqnarray}
\tau^{tot}(m,m_0) &&
\! \! \! \! \! =  2  \int_{m_0}^{m} dy  
  \frac{N  \vert V' (y) \vert}{ p_{eq}(y)}  \int_{-1}^{y} dx p_{eq}(x) 
  =  2  N \int_{m_0}^{m} dy      \vert V' (y) \vert  \int_{-1}^{y} dx e^{ N \left[ V(y)-V(x) \right] }    
  \ \ {\rm for } \ \ m_0 \in [-1,m[
\nonumber \\
\tau^{tot}(m,m_0)  && 
\! \! \! \! \!=  2  \int_{m}^{m_0} dy  
  \frac{N  \vert V' (y) \vert}{ p_{eq}(y)}  \int_{y}^{1} dx p_{eq}(x) 
=
 2 N \int_{m}^{m_0} dy  \vert V' (y) \vert \int_{y}^{1} dx e^{ N \left[ V(y)-V(x) \right] } 
   \ \ {\rm for } \ \  m_0 \in ]m,1]
  \label{tauSDB}
\end{eqnarray}
 should be compared to its detailed-balance counterpart of Eq. \ref{tauDBdiff}.


\subsection{ Equilibrium time $\tau_{eq} $ }

The equilibrium time of Eq. \ref{taueqx0indeo} does not depend on the initial magnetization $m_0$
and can be thus evaluated for the special case $m_0=1$
 \begin{eqnarray}
\tau^{tot}_{eq} && = \tau^{tot}_{eq}(m_0=1) =  \int_{-1}^{+1} dm p_{eq}(m) \tau^{tot}(m,m_0=1) 
 \nonumber \\ &&
 =  \frac{2N }{ \int_{-1}^{+1} dz e^{-  N  V(z) }} \int_{-1}^{+1} dm   e^{-  N  V(m) } 
   \int_{m}^{1} dy  \vert V' (y) \vert \int_{y}^{1} dx e^{ N \left[ V(y)-V(x) \right] } 
\label{taueqSDB}
\end{eqnarray}

For large $N$, in the high-temperature phase and at criticality, it is the expansion around $m=0$
of Eq. \ref{vmserieshc} that will be important. So it is again convenient to use Eq. \ref{vmp}
and to rescale all the integration variables as in Eq. \ref{taueqDBdiffrescal} to obtain for Eq. \ref{taueqSDB}
 \begin{eqnarray}
\tau^{tot}_{eq} \opsimeq_{N \to +\infty} \frac{1}{(\alpha N)^{\frac{1}{p} } } 
 \frac{2 }{ \int_{-\infty}^{+\infty} d{\tilde z} e^{-     {\tilde z} ^p }} 
 \int_{-\infty}^{+\infty} d{\tilde m} e^{-     {\tilde m} ^p }
  \int_{\tilde m}^{+\infty}   d {\tilde y} \ \ \ p \vert   {\tilde y}^{p-1} \vert \int_{\tilde y}^{+\infty}    d{\tilde x}   e^{ {\tilde y} ^p - {\tilde x} ^p  }
\label{taueqSDBrescal}
\end{eqnarray}

In the high-temperature phase with $p=2$ and $\alpha= (1  - \beta J )= (\beta_c-\beta) J$,
the equilibrium time 
 \begin{eqnarray}
\tau^{tot}_{eq} \propto \frac{1}{ \sqrt{ (\beta_c-\beta) N } } \ \ \ \ \ {\rm for } \ \ \beta < \beta_c 
\label{taueqSDBrescalhigh}
\end{eqnarray}
is smaller than its detailed-balance counterpart of Eq. \ref{taueqDBdiffrescalhigh}.
Note however that in the discrete-time formulation where one counts instead the number of Markov-chain steps
\cite{TCV,weigel,vuceljaa,SDBcurierun}, the convergence-time is found to scale as $N^{\frac{1}{2}}$ in the high-temperature phase.
But in both conventions, the ratio of the equilibrium times is given by the system-size scaling as $N^{-\frac{1}{2}} $ recalled in Eq. \ref{ratiotaueqSDB} of the Introduction.

At criticality with $p=4$, the equilibrium time 
 \begin{eqnarray}
\tau^{tot}_{eq} \opsimeq_{N \to +\infty} \frac{1}{N^{\frac{1}{4}} } \ \ \ \ \ {\rm at } \ \ \beta= \beta_c 
\label{taueqSDBdiffrescalcriti}
\end{eqnarray}
is also smaller than its detailed-balance counterpart of Eq. \ref{taueqDBdiffrescalcriti}.
Again in the discrete-time formulation where one counts instead the number of Markov-chain steps
\cite{TCV,weigel,vuceljaa,SDBcurierun}, the convergence-time is found to scale as $N^{\frac{3}{4}}$ at criticality.
But in both conventions, the ratio of the equilibrium times is given by the system-size scaling as $N^{-\frac{3}{4}} $ recalled in Eq. \ref{ratiotaueqSDB} of the Introduction. 

Finally, in the low-temperature phase $\beta>\beta_c$, the leading behavior is the same exponential dependence in $N$
of Eq. \ref{taueqDBdifflow} as for the detailed-balance case.


\end{document}